%% file: ms.tex
\documentclass[twocolumn]{aastex63}
\newcommand{\sftw}[1]{\texttt{#1}}

\usepackage{hyperref}
\usepackage{color}
\usepackage[normalem]{ulem}
\usepackage{amsmath,amssymb}	
\usepackage{mathpazo}
\usepackage{textcomp}
\usepackage{booktabs}
\usepackage{xcolor}
\usepackage{nicefrac}
\usepackage{bm}
\usepackage{graphicx}
\usepackage{url}
\usepackage{savesym}
\usepackage[T1]{fontenc}

\definecolor{green1}{RGB}{0, 128, 0}

\newcommand{\DS}{\displaystyle}

\newcommand{\cG}{{\cal G}}

\newcommand{\cS}{{\cal S}}
\newcommand{\cU}{{\cal U}}

\newcommand{\norm}[1]{ {\left|\left|{#1}\right|\right|}}

\received{May, 2020}
\revised{October, 2020}
\accepted{November, 2020}

\submitjournal{\apj}

\shorttitle{A two-moment Rad-HD scheme applicable to simulations of
planet formation}
\shortauthors{Melon Fuksman, Klahr, Flock and Mignone}

\begin{document}

%\title{A two-moment radiation hydrodynamics scheme applicable to global simulations of protoplanetary disks}
\title{A two-moment radiation hydrodynamics scheme applicable
to simulations of planet formation in 
circumstellar disks}

\correspondingauthor{Julio David Melon Fuksman}
\email{fuksman@mpia.de}

\author[0000-0002-1697-6433]{Julio David Melon Fuksman}
\affil{Max Planck Institute for Astronomy, K\"onigstuhl 17, 69117 Heidelberg, Germany}

\author[0000-0002-8227-5467]{Hubert Klahr}
\affil{Max Planck Institute for Astronomy, K\"onigstuhl 17, 69117 Heidelberg, Germany}

\author[0000-0002-9298-3029]{Mario Flock}
\affil{Max Planck Institute for Astronomy, K\"onigstuhl 17, 69117 Heidelberg, Germany}

\author[0000-0002-8352-6635]{Andrea Mignone}
\affiliation{Dipartimento di Fisica, Universit\`a degli Studi di Torino, via Pietro Giuria 1, 10125 Turin, Italy}

\begin{abstract}

We present a numerical code for
radiation hydrodynamics designed
as a module for the freely available \sftw{PLUTO} code. We adopt a
gray approximation and include
radiative transfer following a
two-moment approach by imposing
the M1 closure to the radiation
fields. This closure allows for
a description of radiative
transport in both the diffusion
and  free-streaming limits, and is
able to describe highly
anisotropic radiation transport as can be expected in the vicinity
of an accreting planet in a protoplanetary disk.
To reduce the computational
cost caused by the timescale
disparity between radiation and
matter fields, we integrate their
evolution equations separately
in an operator-split way, using
substepping to evolve the radiation equations. We further
increase the code's efficiency
by adopting the reduced speed
of light approximation (RSLA).
Our integration scheme for
the evolution
equations of radiation fields
relies on implicit-explicit schemes,
in which radiation-matter interaction
terms are integrated implicitly
while fluxes are integrated
via Godunov-type solvers.
The module is suitable for general astrophysical
computations in $1$, $2$, and $3$
dimensions in Cartesian, spherical
and cylindrical coordinates, and
can be implemented on rotating
frames. We demonstrate the algorithm
performance on different numerical
benchmarks, paying particular attention to the applicability of
the RSLA for computations of physical processes in protoplanetary disks.
We show 2D simulations of vertical
convection in disks and 3D simulations
of gas accretion by planetary cores, which are the first of their
kind to be solved with
a two-moment approach.

\end{abstract}


\input{introduction.tex}

\input{equations.tex}
\input{num_scheme.tex}
\input{application.tex}
\input{summary.tex}

\appendix
\input{performance.tex}

\newpage

\acknowledgements{
We thank Oliver Voelkel and Rolf Kuiper for sharing their manuscript and insight into M1 methods for disk simulations using the \sftw{PLUTO} code.
The research of J.D.M.F.\ and H.K.\ is supported by the German Science Foundation (DFG) under the priority program SPP 1992: "Exoplanet Diversity" under contract KL 1469/16-1. M.F. received funding from the European Research Council (ERC) under the European Union's Horizon 2020 research and innovation program (grant agreement n° 757957). We thank our collaboration partners on this project in Kiel under contract (WO 857/17-1) Sebastian Wolf and Anton Krieger for fruitful discussions and guidance for the synchronization of both project parts. This research was also supported by the Munich Institute for Astro- and Particle Physics (MIAPP) of the DFG cluster of excellence "Origin and Structure of the Universe" and was performed in part at KITP Santa Barbara by the National Science Foundation under Grant No. NSF PHY11-25915.
We also thank the anonymous referee for constructive
comments that helped to improve the quality of this work.
}

\bibliography{refs}

\end{document}

%% file: introduction.tex
 \section{Introduction}\label{S:Introduction}
 
 Radiative transfer a key tool to understand the
 dynamics and observational properties of almost
 any astrophysical system. In protostellar disks, the
 study of radiative processes
 is a necessary ingredient to predict which zones
 are able to develop different hydrodynamical
 instabilities that lead to turbulence and
 consequent transport of angular momentum,
 structure formation, and eventual growth of
 planets
 \citep[see, e.g.,][]{Gammie1996,Flock2017InnerRim,Manger2018,Pfeil2019}.
 Some processes, such as diffusive
 cooling or the radiative processes occurring in
 the vicinity of gap-opening planets, 
 may involve transport of radiation between
 optically thick and optically thin regions. 
 This can in principle lead to anisotropic
 transport regimes involving highly beamed
 radiative intensities, which require a proper
 treatment that allows for such directional transport.
 On the other hand, a self-consistent treatment of
 stellar irradiation and dust absorption,
 emission, and scattering of radiation
 coupled to gas dynamics is needed to explain the current observations of
 disk substructures at increasingly high resolution in the thermal dust emission \citep[see, e.g.,][]{ALMA2015,Flock2015}.
 
 The coupled integration of hydrodynamics (HD) and
 frequency-dependent radiative transfer is in general
 a computationally expensive task,
 and approximate
 methods are most usually preferred.
 A generally adopted assumption is the gray approximation,
 in which the radiative intensity and the material
 absorption and scattering opacity coefficients
 are averaged in
 the frequency domain. This approach leads to a
 description of total energy and momentum exchange
 between matter and radiation,
 without regarding
 frequency-dependent phenomena. The applicability
 of the gray approximation is tied to the variation of the
 material's opacity with frequency in the spectral
 region of interest, and is therefore case-dependent.

Among all gray radiative transfer schemes, the flux-limited diffusion (FLD) method by
 \cite{Levermore1981FLD}
 is the most widely preferred method
 in the context of protoplanetary
 disks and star formation in general. This
 is a one-moment method, meaning that the full
 radiative transfer equation is turned into
 a single evolution equation for one of the
 moments (angular integrals) of the 
 specific radiative intensity, in this case, the radiation
 energy density. In FLD, the radiation flux
 is computed via an ad hoc function of the
 radiation energy density, its gradient, and
 the material's local opacity, in such a way
 that the module of the flux tends to its
 correct limit in the diffusion and
 free-streaming regimes. This method is
 particularly accurate in highly opaque systems,
 where the radiation transport equation correctly
 tends to a diffusion equation. Conversely, due to the adopted 
 definition
 of the radiation flux, some degree of inaccuracy is generally observed
 in regions of low opacity 
 \citep{Rosdahl2015}.
 On the other hand,
 FLD methods are unable to describe
 strongly anisotropic transport
 in phenomena
 such as shadows or simply free streaming, in which
 cases they introduce unphysical numerical
 diffusion due to the fact that the radiation flux
 is always proportional to the gradient of the
 energy density \citep{Hayes2003}.
 
To make predictions on the observational appearance of accreting planets \citep{2018MNRAS.473.3573S, 2019MNRAS.487.1248S}
and to reconstruct the characteristics of exoplanets from observations of disks around young stars, one needs a combination of radiation hydrodynamical simulations in the gray approximation, as we can provide in this paper,
and detailed Monte Carlo continuum radiative transfer simulations, as presented by our collaboration partners \citet{Krieger2020}. In
subsequent works, we intend to connect in this way realistic flow and temperature structures with frequency-dependent intensity maps for various instruments such as ALMA \citep{2002Msngr.107....7K}, PIONIER \citep{2011A&A...535A..67L}, and MATISSE \citep{2014Msngr.157....5L}.
On the other hand, the growth time scale of gas planets \citep{2012A&A...547A.111M} depends on the efficiency of radiative cooling \citep{2013ApJ...778...77D, 2014ApJ...782...65S,2016MNRAS.460.2853S,2017MNRAS.465L..64S,Schulik2020} and therefore a better understanding of possible gas accretion rates also in the presence of pebble \citep{2006ApJ...639..432K, 2010A&A...520A..43O, 2012A&A...544A..32L} and planetesimal accretion \citep{Fortier2013} will have a strong impact on the ability to form efficiently gas giants. 
Due to the mentioned low opacity
regimes occurring, e.g., in planetary
gaps, it is ideal to count with 
radiation transport schemes
that do not rely on a pure diffusion 
approximation.
 
 In this work we have implemented the two-moment approach by \cite{Levermore1984M1}, generally referred to as M1 closure. In this method, an additional set of equations is solved for the radiation flux
 components, where this time the radiation
 pressure tensor is defined in terms of the
 radiation flux and energy density. This
 closure is based on the assumption 
 that the specific radiative intensity is isotropic
 in a given reference frame, and hence it yields
 exact flux values if such assumption is correct.
 Despite this is often a fairly reasonable
 approximation, it must be noted that this
 assumption fails to describe cases where
 such a reference frame does not exist. This happens,
 for instance, when optically thin regions of space
 have converging beams that originate
 from different directions, in which case the M1 closure produces unphysical interactions between
 the beams \citep[see, e.g.,][]{Sadowski2013, Skinner2013}.
 Another important advantage of this closure is that
 freely streaming radiation fields are transported maintaining their original direction, without being artificially spread as in FLD methods.
 On the other hand, both
 methods yield the same diffusion equation in largely
 opaque media.
 From the numerical point of view, the M1 closure
 counts with the advantage that the evolution
 equations are hyperbolic with local interaction
 source terms, whereas the FLD equations are
 parabolic and usually solved via fully implicit
 methods \citep[see][]{Commercon2011}. Hence, unless fully implicit schemes are used to solve the evolution equations,
 M1 methods should have favorable
 scaling properties when compared to FLD.
 
 We have implemented a two-moment radiation HD
 (Rad-HD) module within the multi-algorithm,
 high-resolution code \sftw{PLUTO}, designed for time-dependent computations of relativistic or nonrelativistic unmagnetized or magnetized flows
 \citep{Mignone2007}.
 The module is fully parallel,
 and can be applied using
 Cartesian, cylindrical, and
 spherical coordinate
 systems in $1$, $2$ or $3$ dimensions.
 Our current implementation is an
 extension of
 the module for
 radiation relativistic
 magnetohydrodynamics
 (Rad-RMHD) introduced in \cite{MelonFuksman2019}, where implicit-explicit
 (IMEX) schemes have been used to integrate the
 evolution equations in such a way that fluxes are integrated explicitly, while the 
 potentially stiff radiation-matter interaction
 terms are integrated implicitly.
 In that case, the time step is
 computed as a minimum of the
 maximum time steps allowed for
 the transport of radiation and magnetohydrodynamical fields,
 obtained in each case
 by applying the Courant-Friedrichs-Lewy (CFL) stability condition \citep[][]{Courant1928}.
 Contrarily, in our case,
 radiation and nonrelativistic flows
 evolve in largely different timescales,
 which renders that approach
 computationally prohibitive
 and largely diffusive due to
 the accumulation of truncation
 error. To reduce the computational
 cost, we follow a twofold strategy.
 On the one hand, we adopt the reduced
 speed of light approximation (RSLA), introduced by \cite{GnedinAbel2001}
 and applied to M1 Rad-HD by \cite{Skinner2013},
 in which the value of the speed of
 light is replaced by an artificially
 low value in order to reduce
 the mentioned scale disparity.
 This increases the maximum time step
 allowed by the CFL condition,
 consequently reducing the overall
 cost of the operations.
 The RSLA is valid as long
 as the chosen reduced value of the
 speed of light is larger
 than any velocity scale in the
 problem at hand, in which case it
 yields the same solutions that
 would be obtained using
 its physical value.
 Since this restriction maintains
 some disparity between the mentioned timescales, we further reduce the computational cost of the method by applying operator splitting to solve the HD and radiation equations in different
 steps. We use in each case the 
 corresponding time step restriction
 given by the CFL condition and apply
 substepping to solve the radiation
 subsystem, using IMEX schemes
 to integrate the radiation fields.
 
 %Additional features of the code include an adaptation of the Harten-Lax-van Leer–contact (HLLC) solver for radiation transport introduced in \cite{MelonFuksman2019}, and the ability {\bf to solve} the Rad-HD equations in a rotating frame following the conservative formulation described in \cite{Mignone2012FARGO}.
% {\bf
% Together with the second-order
% accuracy achieved by the
% implemented IMEX-SSP2(2,2,2)
% scheme \citep{PareschiRusso2005},
% also implemented in the general
% relativistic code by \cite{McKinney2014},
% these are improvements with respect to other existing M1 methods, such as that introduced by \cite{Skinner2013}. Furthermore,
% the numerical diffusion
% introduced by the operator-split radiative transfer scheme
% applied in that work
% causes shadow profiles to be
% appreciably less defined than
% those obtained with the
% IMEX schemes implemented in
% our module 
% \citep[see][]{MelonFuksman2019}.
% Even though it is our particular
% interest to apply our module to
% planet formation scenarios, its
% applicability is rather general,
% and it will be included in future versions of \sftw{PLUTO}.
 
 Additional features of the code include an adaptation of the Harten-Lax-van Leer–contact (HLLC) solver for radiation transport introduced
 in \cite{MelonFuksman2019}
 and an implementation of the 
 second-order accurate IMEX-SSP2(2,2,2)
 scheme by \cite{PareschiRusso2005},
 also implemented in the general
 relativistic code by \cite{McKinney2014},
 both of which represent
 improvements with
 respect to other existing M1 methods, such as that introduced by \cite{Skinner2013}.
 Furthermore,
 the numerical diffusion
 introduced by the operator-split radiative transfer scheme
 applied in that work
 causes shadow profiles to be
 appreciably less defined than
 those obtained with the
 IMEX schemes implemented in
 our module 
 \citep[see][]{MelonFuksman2019}.
 On the other hand, the code can be applied to solve the Rad-HD equations
 in a rotating frame following the conservative formulation described in \cite{Mignone2012FARGO}, under the
 condition that the relativistic 
 corrections appearing when
 transforming the radiative transport equations into such frame can be disregarded.
 This feature is particularly 
 useful in planet formation scenarios to limit the numerical diffusion in the vicinity of accreting planets.
 Even though it is our particular interest to apply our module to such systems, its applicability is rather general, and it will be included in future versions of \sftw{PLUTO}.
 % Even though such systems are the main driver for the development of our module, its applicability is rather general, and it will be included in future versions of \sftw{PLUTO}.
 
 Several two-moment Rad-HD implementations can be found
 in the literature \citep[see, e.g.,][]{Audit2002,Hayes2003,Gonzalez2007,JiangStone2012, Sadowski2013,Skinner2013,Takahashi2013,McKinney2014,Rosdahl2015,MelonFuksman2019,Weih2020,MignonRisse2020}.
 To our knowledge, these methods
 have not been yet applied to model protoplanetary disk evolution and planet formation scenarios, besides in a submitted paper by (Voelkel and Kuiper, A$\&$A, submitted). Note that these authors implemented a fully implicit scheme, which does not make use of the reduced speed of light ansatz, yet makes global parallelisation and adaptive mesh refinement less efficient. 
 An interesting application in the context of star formation is shown in \cite{MignonRisse2020}, where the formation of a disk following the collapse of a massive prestellar
 core is studied using a hybrid method in which stellar irradiation is modelled with an M1 scheme, while gas reemission and absorption is treated via FLD. In this work, we have studied different applications of our module to global simulations of protoplanetary disks, paying special attention to the applicability of the RSLA in this context. In particular,
 we have modelled the growth of the vertical convective instability in
 a disk and the accretion of gas onto a planetary core.

 This paper is organized as follows. In Section \ref{S:Equations}, we
 summarize the main equations characterizing our
 model and discuss
 the main features and
 limitations of the RSLA, while in Section \ref{S:NumScheme} we describe the implemented 
 algorithms. In Section
 \ref{S:Applications}, we test the code's performance on different numerical benchmarks and study
 different applications in the
 context of protoplanetary disks.
 In Section \ref{S:Conclusions}, we summarize the
 main results of our work.
 Additional performance tests
 and comparisons to other methods
 are included in Appendix \ref{S:Performance}.

%% file: equations.tex
 \section{Governing equations}\label{S:Equations}

 \subsection{Radiation hydrodynamics}\label{SS:RadHD}
Throughout this work we solve the equations of a fluid interacting with
a radiation field, for which we follow
a two-moment approach under the gray approximation.
The resulting evolution equations, namely the Rad-HD equations, 
can be written in quasi-conservative form as
\begin{equation}\label{Eq:RadHD}
\begin{split}
\frac{\partial \rho}{\partial t} + \nabla \cdot 
\left(\rho \mathbf{v}\right) &= 0 \\
\frac{\partial( \rho \mathbf{v})}{\partial t} + \nabla \cdot 
\left(\rho \mathbf{v} \mathbf{v}\right)+\nabla p_g &= 
   \mathbf{G}    + \mathbf{S}_\mathbf{m} -\rho\nabla \Phi  \\
\frac{\partial\left( E+\rho\Phi\right)}{\partial t} + \nabla \cdot 
\left[(E+p_g+\rho\Phi) \mathbf{v}\right] &= c\,G^0  +S_\mathrm{E}
  -\nabla\cdot\mathbf{F}_\mathrm{Irr} \\
\frac{1}{\hat{c}}\frac{\partial E_r}{\partial t}+\nabla\cdot \mathbf{F}_r &= -G^0 \\
\frac{1}{\hat{c}}\frac{\partial \mathbf{F}_r}{\partial t}+\nabla\cdot \mathbb{P}_r &= -\mathbf{G}\,,
\end{split}
\end{equation}
where $\rho$, $p_g$, and $\mathbf{v}$ are the fluid's density,
pressure and velocity, while $E_r$, $\mathbf{F}_r$, and
$\mathbb{P}_r$ are respectively the radiation energy, flux, and
pressure tensor. The gas energy density $E$ is defined in terms
of these fields as
\begin{equation}\label{Eq:GasEnergyDensity}
E = \rho \epsilon + \frac{1}{2}\rho \mathbf{v}^2\,,
\end{equation}
where $\rho\epsilon$ is the gas internal energy density. On the
other hand, radiation fields are defined in terms of the frequency-
and direction-dependent radiation specific intensity $I_\nu(t,\mathbf{x},\mathbf{n})$, as
\begin{equation} \label{Eq:RadMomentsF}
\begin{split}
    E_r &\DS = \frac{1}{c} \int_0^\infty\mathrm{d}\nu 
					 \oint  \mathrm{d}\Omega\,\,
  					 I_\nu(t,\mathbf{x},\mathbf{n}) \\
    F_r^i&\DS = \frac{1}{c} \int_0^\infty\mathrm{d}\nu 
					 \oint  \mathrm{d}\Omega\,\,
  					 I_\nu(t,\mathbf{x},\mathbf{n})\, n^i \\
    P_r^{ij} &\DS = \frac{1}{c} \int_0^\infty\mathrm{d}\nu 
					 \oint  \mathrm{d}\Omega\,\,
  					 I_\nu(t,\mathbf{x},\mathbf{n})\, n^i\,n^j 	 
\end{split}
\end{equation}
\citep[see][]{Mihalas}, in such a way that all three quantities
are measured in units of energy density. Additionally, we have
included a gravitational potential $\Phi$, which is defined as a general function of the spatial coordinates. The constants $c$
and $\hat{c}$ correspond, respectively, to the speed of light
and its reduced value (see Section \ref{SS:RSLA}). 
In our implementation,
these equations can
be solved in Cartesian, cylindrical, or spherical coordinates.

Several source terms are included on the right-hand side of
Eq. \eqref{Eq:RadHD},
beginning with the radiation-matter interaction terms $G^0$
and $\mathbf{G}$. In the gray approximation, these can be written
in the fluid's comoving frame as 
\begin{equation}\label{Eq:Gcomov}
  \begin{split}
  	\tilde{G}^0&=\kappa\,\rho\left(\tilde{E}_r- a_R T^4\right) \\
  	\tilde{\mathbf{G}}&=\chi\,\rho\,\tilde{\mathbf{F}_r}\,,
  \end{split}
\end{equation}
where $a_R = \sigma_\mathrm{SB}/\pi c$ is the radiation constant,
$\sigma_\mathrm{SB}$ the Stefan-Boltzmann constant,
T the gas temperature,
and $\kappa$, $\sigma$, and $\chi=\kappa+\sigma$ are, respectively, the
frequency-averaged absorption, scattering, and total opacity
coefficients, which can be defined as general functions of $\rho$ and $T$.
It is customary to compute $\kappa$
and $\chi$ in Eq. \eqref{Eq:Gcomov},
respectively, as their Planck and Rosseland means, since
the first of these choices
is particularly accurate for
low opacities while the second
one yields the correct flux
in the diffusion regime
\citep{Mihalas}. For testing purposes, unless otherwise stated,
we take these averages to be equal, and use single values for $\kappa$, $\sigma$, and $\chi$ keeping in
mind that the actual values
can be largely different
when different averaging
procedures are applied
\citep[see, e.g.,][]{Malygin2014}. 
Opacity coefficients,
together with quantities
under tilde, are measured in the comoving frame, whereas
every other quantity is measured in the laboratory frame.
Gas temperatures are computed following the ideal law
\begin{equation}\label{Eq:TempIdealGas}
T = \frac{\mu u}{k_\mathrm{B}}\frac{p_g}{\rho}\,,
\end{equation}
where $\mu$ is the gas mean molecular weight, $u$ is the
atomic mass unit, and $k_\mathrm{B}$ is the Boltzmann constant. We compute
the interaction terms in the laboratory frame by 
making use of the following Lorentz transformation laws
to first order in $\bm\beta=\mathbf{v}/c$:
\begin{equation}\label{Eq:BoostG}
  \begin{split}
  G^0 &= \tilde{G}^0 + \bm{\beta}\cdot\tilde{\mathbf{G}}\\
  \mathbf{G} &= \tilde{\mathbf{G}} +\tilde{G}^0 \bm{\beta} \,.
  \end{split}
\end{equation}
Similarly, the radiation fields are transformed into the laboratory
frame to first order in $\beta$, as
  \begin{equation}\label{Eq:TransformationLaws}
  \begin{split}
  E_r &= \tilde{E}_r + 2 \beta_i \tilde{F}_i\\
  F_r^i&=\tilde{F}_r^i+\beta^i\tilde{E}_r +\beta_j\tilde{P}^{ij}_r\\
  P_r^{ij}&=\tilde{P}^{ij}_r + \beta^i \tilde{F}^j + \beta^j \tilde{F}^i   \,.
  \end{split}
  \end{equation}
This yields the following expressions for the interaction terms
that are used in the code:
  \begin{equation}\label{Eq:SourceTerms}
  \begin{split}
  G^0 &= \rho \kappa \left( E_r - a_R T^4 -
  2 \bm\beta \cdot \mathbf{F}_r \right)\\
  &+\rho\chi\,\bm\beta\cdot\left(
		\mathbf{F}_r-E_r\bm\beta-\bm\beta \cdot\mathbb{P}_r  
  \right)\\
  \mathbf{G} &= \rho \kappa \left( E_r - a_R T^4 -
  2 \bm\beta \cdot \mathbf{F}_r \right)\bm\beta \\
  &+\rho\chi\bm\left(
		\mathbf{F}_r-E_r\bm\beta-\bm\beta \cdot\mathbb{P}_r  
  \right)\,,
  \end{split}
  \end{equation}
  where we have kept some $\mathcal{O}(\beta^2)$ terms
  in order to recover the local thermal equilibrium (LTE) limit given by 
   $\tilde{E}_r\rightarrow  a_R T^4$ and
   $\tilde{F}_r\rightarrow\mathbf{0}$ 
   when $\sigma,\kappa\rightarrow\infty$
   \citep[similar approaches are followed in][]{LowrieMorel1999,JiangStone2012}.
  
  An irradiation term \mbox{$-\nabla\cdot\mathbf{F}_\mathrm{Irr}$}
  is included in Eq. \eqref{Eq:RadHD} to account for radiative heating
  caused by sources emitting in a different frequency range than
  the one considered in the radiation transport scheme.
  One such example is the heating from star irradiation in
  protoplanetary disks, in which the radiation coming
  from the star peaks in the visible range, but most of the 
  energy emitted by the dust is in the infrared.
  This additional flux is not updated by solving an evolution
  equation, but it is instead computed at each time step as a
  function of space.   
  
  Finally, the terms $S_E$ and $\mathbf{S}_m$
  account for dissipative effects included in the current
version of \sftw{PLUTO}, such as thermal conduction, optically thin cooling, and viscosity \citep{Mignone2012}.
In the latter case, these terms take the form
\begin{equation}\label{Eq:ViscSourceTerms}
\begin{split}
\mathbf{S}_\mathbf{m} &= \nabla \cdot \Pi \\
S_E &= \nabla \cdot \left(\mathbf{v}\cdot\Pi\right) \,,
\end{split}
\end{equation}
where $\Pi$ is the viscosity tensor defined as
\begin{equation}\label{Eq:ViscTensor}
\Pi = \rho \nu_1 \big[ \nabla \mathbf{v} +
\left( \nabla \mathbf{v}\right)^\intercal \big] 
+ \rho\left( \nu_2 - \frac{2}{3} \nu_1\right)
\left(\nabla\cdot\mathbf{v}\right) \mathbb{I} \,,
\end{equation}
where $\mathbb{I}$ is the identity matrix,
while $\nu_1$ and $\nu_2$ are, respectively, the shear and bulk viscosity coefficients.

In cylindrical and spherical
coordinates,
Eq. \eqref{Eq:RadHD} can
be integrated in a reference
frame that rotates with a
uniform angular velocity $\Omega$.
The integration of the additional terms that appear when
applying Galilean transformations to the HD fields
follows the conservative formulation detailed
in \cite{Mignone2012FARGO}.
On the other hand,
$E_r$ and $\mathbf{F}_r$ follow the transformation law
given by Eq. \eqref{Eq:TransformationLaws}, and therefore all additional terms
arising from this transformation are of order $\Omega R/c$,
where $R$ is the cylindrical radius. In the current form of the module we do not include such
additional terms, which means that
the rotating frame scheme
can only be applied when terms of order $\bm\beta$ can be disregarded,
as is typically the case in
planet formation scenarios
(see, e.g., Section \ref{SS:Planet}).
This means that, in such cases, the radiation-matter interaction terms are equal to their comoving
values (Eq. \eqref{Eq:Gcomov}).
However, since the relativistic
corrections to the HD equations
are of order $\bm\beta^2$, we keep
in general all terms of order 
$\bm\beta$ in Eq. 
\eqref{Eq:SourceTerms}
to account for mildly relativistic cases where $\bm\beta^2$
can be disregarded.

 \subsection{Closure relations}\label{SS:Closure}
 
The system of equations \eqref{Eq:RadHD} is completely defined by imposing a series of closure relations.
For HD quantities, we impose the equation of state of an ideal gas,
\begin{equation}\label{Eq:EoS}
    \rho \epsilon = \frac{p_g}{\Gamma - 1}\,,
\end{equation}
with a constant specific heat ratio $\Gamma$. For the radiation
fields, we implement the M1
closure \citep{Levermore1984M1},
in which
the components of the pressure
tensor can be computed in terms of $E_r$ and $\mathbf{F}_r$ as
  \begin{equation}\label{Eq:M11}
  P_r^{ij}=D^{ij}E_r\,,
  \end{equation}
  where the Eddington tensor is defined as
  \begin{equation}
  D^{ij}=\frac{1-\xi}{2}\,\delta^{ij}+
  \frac{3\xi-1}{2}n^in^j\,,
  \end{equation}
  with
  \begin{equation}\label{Eq:M13}
  \xi=\frac{3+4f^2}{5+2\sqrt{4-3f^2}}\,,
  \end{equation}
  where $\bm{n}=\mathbf{F}_r/\vert\vert\mathbf{F}_r\vert\vert$,
  $f=\vert\vert\mathbf{F}_r\vert\vert/E_r$, and
  $\delta^{ij}$ is the Kronecker delta. With these definitions,
  the radiation fields correctly reproduce both the
  free-streaming limit when
  $\vert\vert\mathbf{F}_r\vert\vert\rightarrow E_r$, in which
  case $P_r^{ij}=E_r\,n^in^j$, and the diffusion limit when
  $\vert\vert \mathbf{F}_r \vert\vert \ll E_r$, which gives 
  the Eddington approximation $P_r^{ij}=\left(\delta^{ij} \middle/ 3\right)E_r$. The latter case is verified for 
  large opacities, in which case the last two of equations
  \eqref{Eq:RadHD} yield the diffusion equation
  \begin{equation}\label{Eq:Diff}
      \frac{\partial E_r}{\partial t} \approx 
      \nabla\cdot \left(
      \frac{\hat{c}}{3\rho\chi} \nabla E_r
      \right) 
  \end{equation}
  for slow variations of $\partial_t\mathbf{F}_r$.
  Equation \eqref{Eq:Diff} shows that the diffusion 
  coefficient has been artificially reduced by a
  factor $c/\hat{c}$, which limits the applicability
  of this method to cases that are at most weakly
  dependent on its physical value, as detailed in
  Section \ref{SS:RSLA}.

 \subsection{The reduced speed of light approximation}\label{SS:RSLA}

 The RSLA consists in choosing
 a value of $\hat{c}$
  smaller than $c$, in such a way to reduce the computational cost of integrating Eq. \eqref{Eq:RadHD} (see Section \ref{S:NumScheme}). 
  This formalism has the drawback of introducing
  unphysical phenomena, the most evident one being
 that the propagation velocity of freely streaming radiation
 fields is $\hat{c}$ instead of $c$. On the other hand,
 radiation-matter interaction timescales
 such as thermal equilibrium and diffusion timescales are increased 
 (see, e.g., Eq. \eqref{Eq:Diff}). Another important consequence
 of this approach is that the usual form
 of the conservation of
 total energy-momentum is lost.
 Disregarding gravity
 and all nonideal source terms in Eq. \eqref{Eq:RadHD} except
 for the radiation-matter interaction terms, we can obtain
 conservation laws for the fields 
 \begin{equation}\label{Eq:Etotmtot}
 \begin{split}
     E_\mathrm{tot} &=E+(c\,/\hat{c})E_r \\
     \mathbf{m}_\mathrm{tot} &=\rho\mathbf{v}+
     (1/\hat{c})\mathbf{F}_r\,,
\end{split}
 \end{equation}
 which are only equal to the total energy and momentum densities
 if $\hat{c}=c$. Still, the RSLA yields exact stationary
 solutions of Equations \eqref{Eq:RadHD},
 since $\hat{c}$ does not appear in them
 if all time derivatives are set to zero.
 More generally, the RSLA yields exact solutions of the Rad-HD
 equations provided radiation-matter interaction occurs
 much faster than
 any timescale of interest in the problem at hand.

 A rather general criterion for the applicability of the RSLA
 has been derived in \cite{Skinner2013}, by requiring
 that the existing timescale hierarchies remain
 unchanged when $c$ is replaced
 by $\hat{c}$.
 This condition is satisfied under the conditions that
 the value of $\hat{c}$ remains much larger than the maximum fluid velocity $v_\mathrm{max}$
 and that the diffusion timescale $t_\mathrm{diff}=L\tau_\mathrm{max}/\hat{c}$ is much smaller than the dynamical timescale $t_\mathrm{dyn}=L/v_\mathrm{max}$,
 where $L$ and $\tau_\mathrm{max}$ are a typical length and optical depth of the system.
 These constraints can be summarized as
\begin{equation}\label{Eq:ConstraintC}
\hat{c} \gg v_\mathrm{max}\, \mathrm{max}\left( 1, \tau_\mathrm{max}\right)\,.
\end{equation}
However, it must be
noted that this is an approximate relation,
and that the
determination of an
optimal $\hat{c}$ value
depends in general
on the problem at hand, and can only
be safely achieved through careful testing (see, e.g., Section \ref{SS:Shocks}).

 Using Eq. \eqref{Eq:ConstraintC}, we can
 evaluate the applicability of 
 the RSLA to simulations of
 protoplanetary disks. To this purpose, we consider a disk
 with a superficial density $\Sigma^\mathrm{dust}_{1\,\mathrm{AU}}\sim 10$ g cm$^{-2}$
 undergoing changes that propagate at $v_\mathrm{max}\sim c_s/10$, where $c_s\sim 1$ km s$^{-1}$. Assuming a vertical Gaussian profile at a radius $R=1$ AU
 with $H/R=0.05$, where $H$ is the
 pressure scale height,
 we have computed the 
 optical depth in the vertical direction in an inner zone 
 given by $z\in [ -H,H]$ and an outer zone given by $z\in[ H,4H]$,
 using a mean absorption opacity $\kappa\sim 400$ cm$^2$ g$^{-1}$. 
 This gives the conditions  $\hat{c}\gg c/1100$ for the inner zone and $\hat{c}\gg  c/4700$ for the outer zone.
 The effect of the choice of $\hat{c}$ in this context is further studied in Section \ref{S:Applications}.

%% file: num_scheme.tex
 \section{Numerical scheme}\label{S:NumScheme}
 \subsection{Outline of the algorithm}\label{SS:OutlineAlgorithm} 
 Our integration scheme consists
 of two main steps: the HD step, given by
 the integration of the subsystem
 \begin{equation}\label{Eq:HDStep}
     \frac{\partial \cU_\mathrm{HD}}{\partial t}
     + \nabla \cdot \Phi_\mathrm{HD} 
     = \cS_\mathrm{HD}\,,
 \end{equation}
 and a radiation step, consisting of the 
 integration of the radiation transport and
 interaction terms as
 \begin{equation}\label{Eq:RadStep}
     \begin{split}
        \frac{1}{\hat{c}}
        \frac{\partial \cU_\mathrm{r}}{\partial t}
         + \nabla\cdot \Phi_r &= -\cG \\
         \frac{\partial}{\partial t}
         \left(
        \begin{array}{c}
        E/c\\
        \rho\mathbf{v}
        \end{array} 
    \right)
    &= \cG\,,
     \end{split}
 \end{equation}
 where $\cU_\mathrm{HD} = \left(
 \rho,\,\rho \mathbf{v},\, E + \rho \Phi
 \right)^\intercal$ and
 $\cU_r = \left(E_r,\,\mathbf{F}_r\right)^\intercal$
 are, respectively, the HD and radiation
 conserved fields, 
 $\Phi_\mathrm{HD}=\left(
 \rho\mathbf{v},\,
 \rho \mathbf{v}\mathbf{v} + p_g \mathbb{I},\,
 (E + p_g+ \rho \Phi)\mathbf{v}
 \right)^\intercal$ and 
 $\Phi_r=\left(
 \mathbf{F}_r,\,\mathbb{P}_r
 \right)^\intercal$ are the HD and radiation fluxes,
 and the source terms are defined as
 $\cS_\mathrm{HD} = \left(
 0,\, \mathbf{S}_\mathbf{m},\,
  S_E - \nabla\cdot\mathbf{F}_\mathrm{Irr}
 \right)^\intercal$ and
 $\cG = \left(
 G^0,\mathbf{G}
 \right)^\intercal$.
 
 Following the second-order operator splitting
 scheme by \cite{Strang1968}, our algorithm
 is divided in three consecutive integration steps,
 beginning by a radiation step with a time increment
 $\Delta t = \Delta t^n_\mathrm{HD}/2$,
 followed by a HD step with $\Delta t = \Delta t^n_\mathrm{HD}$ and a final radiation step with
 $\Delta t = \Delta t^n_\mathrm{HD}/2$. 
 For each time step $n$, the time increment $\Delta t^n_\mathrm{HD}$ is updated applying the CFL condition
 to the subsystem given by Eq. \eqref{Eq:HDStep},
 implemented as
 \begin{equation}\label{Eq:CourantHD}
  \Delta t_\mathrm{HD}^{n+1}= C_a
  \min_{ijk}
  \left[
  \frac{1}{N_\mathrm{dim}}
  \sum_d
  \frac{\lambda_{\mathrm{HD}}^d}{\Delta l^d}
  \right]^{-1}\,,
 \end{equation}
 where $\Delta l^d$ and $\lambda_{\mathrm{HD}}^d$ are
 the cell width and maximum signal speed of the HD subsystem 
 along the direction $d$
 at the
 position $(i,j,k)$, while
 $C_a$ is the Courant factor
 and $N_\mathrm{dim}$ is the number of dimensions. Each radiation step is divided into $N_r$ integration substeps,
 where the time increments
 are updated as
 \begin{equation}\label{Eq:CourantRad}
  \Delta t_\mathrm{r}^{n,q+1}= 
  \min \left[C_a
  \min_{ijk}
  \left[
   \frac{1}{N_\mathrm{dim}}
  \sum_d
  \frac{\lambda_{r}^d}{\Delta l^d}
  \right]^{-1},
  \frac{\Delta t_\mathrm{HD}^{n}}{2}-
  \sum_{s=1}^{q} \Delta t_r^{n,s}
  \right]\,,
 \end{equation}
 in such a way that they verify
 the CFL condition and also satisfy
 \begin{equation}
     \sum_{q=1}^{N_r} \Delta t_r^{n,q}
     = \frac{\Delta t_\mathrm{HD}^{n}}{2}\,,
 \end{equation}
 where now $\lambda_r^d$ is the
 maximum signal speed of the radiation subsystem
 (Eq. \eqref{Eq:RadStep}), typically of the same order
 of magnitude as $\hat{c}$.
 This method, similar to that
 applied in \cite{Skinner2013}, reduces
the computational overhead of the HD step if compared to an IMEX scheme applied to the full system of Rad-HD equations, since the radiation and HD time steps generally satisfy
 $\Delta t_r^{n,q}\ll \Delta t_\mathrm{HD}^{n}$
 (see Eqs. \eqref{Eq:ConstraintC}, \eqref{Eq:CourantHD}, and
 \eqref{Eq:CourantRad}). 
 We describe the integration methods implemented in the HD and radiation steps in Sections
 \ref{SS:HDStep} and \ref{SS:RadStep}.
 
 \subsection{HD step}\label{SS:HDStep} 

 Except for the irradiation term, Equation
 \eqref{Eq:HDStep} contains the system of equations
 solved by the HD module of \sftw{PLUTO}, and hence
 its integration scheme remains unchanged with respect to that implemented in the code
 \citep[see][]{Mignone2007}. We
 follow a finite volume approach, in which
 the cell-averaged values of the conserved fields are explicitly integrated by means of total variation diminishing (TVD) Runge-Kutta schemes \citep{GottliebShu1996}, making use of Godunov-type
 solvers to compute fluxes at zone interfaces.
 To this purpose, volume averages are reconstructed 
 at cell boundaries using piecewise monotonic interpolants inside each computational cell.
 
 During the HD step, all source terms are computed
 at cell centers and explicitly integrated together
 with flux divergences. If irradiation is implemented, the value of
 $\mathbf{F}_\mathrm{Irr}$ is updated at each
 time step according to the current
 mass distribution,
 and its divergence is stored at cell
 centers for its integration.
 On the other hand, parabolic source terms
 such as the viscosity terms in Eqs.
 \eqref{Eq:ViscSourceTerms} can be either
 explicitly integrated in a single time
 step or in several substeps by means of
 one of the super-time-stepping (STS)
 techniques introduced in \cite{Alexiades1996}
 and \cite{Meyer2012}, whose implementation in
 \sftw{PLUTO} is discussed in 
 \cite{Mignone2007} and \cite{Vaidya2017}.
 If STS is used, the HD time increment is
 computed following Eq. \eqref{Eq:CourantHD},
 otherwise being reduced following the
 prescription by \cite{Beckers1992}
 to account for additional stability
 conditions for the integration of parabolic terms.
 
 \subsection{Radiation step}\label{SS:RadStep} 
 The methods followed during the radiation
 step are based on those implemented in
 in \cite{MelonFuksman2019}. In this work, Equation \eqref{Eq:RadStep} is integrated by means of IMEX-Runge Kutta schemes, which consist of
 modified
 Runge-Kutta schemes in which all fluxes are integrated explicitly, while radiation-matter
 interaction terms are integrated implicitly. In
 particular, we have
 implemented the IMEX-SSP2(2,2,2) method
 by \cite{PareschiRusso2005},
 and the IMEX1 method employed in
 \cite{MelonFuksman2019},
 also implemented by \cite{BucciantiniDelZanna2013} in the context
 of resistive general relativistic
 magnetohydrodynamics. These methods
 are of order 2 and 1 in time and L- and A-stable, respectively, which makes 
 IMEX-SSP2(2,2,2) a more robust option
 in some applications, being stable for
 larger values of $C_a$. On the other hand,
 IMEX1 seems to be a more accurate option able to balance out
 advection and interaction terms
 in problems where both are much
 larger than their difference,
 as is the case in diffusion problems \citep{MelonFuksman2019}.
 Both methods are further
 compared in Section
 \ref{SS:LinearWaves},
 where they are used to compute
 the evolution of damped linear radiation waves.
 
 During each explicit step of the mentioned IMEX schemes, an equation of the form
 \begin{equation}
     \frac{1}{\hat{c}}
        \frac{\partial \cU_\mathrm{r}}{\partial t}
         + \nabla\cdot \Phi_r = 0
 \end{equation}
 is explicitly integrated by applying
 a TVD Runge-Kutta scheme and using
 Godunov-type solvers to compute fluxes
 at zone interfaces, as done in the HD step.
 We have implemented three different Riemann
 solvers: a Lax-Friedrichs-Rusanov solver
 \citep[see, e.g.,][]{Toro}, the
 Harten-Lax-van Leer (HLL) solver by
 \cite{Gonzalez2007}, and the
 HLLC
 solver introduced in \cite{MelonFuksman2019}.
 Characteristic radiation velocities
 are computed as described in \cite{Audit2002}
 and \cite{Skinner2013}, and limited
 in optically thick cells in order
 to minimize numerical diffusion according to
 the prescription introduced in
 \cite{Sadowski2013}. The upper limit to
 the radiation flux given by the physical
 constraint
 \begin{equation}
     \norm{\mathbf{F}_r} \leq E_r
 \end{equation}
 is imposed on cell boundaries during the reconstruction step. On the other hand, geometrical source terms that arise from the expression of the divergence in curvilinear coordinates are explicitly integrated during the explicit step.
 
 All remaining terms in Equation \eqref{Eq:RadStep} are integrated
 in the implicit step. To do so, we
 rearrange this equation in the
 following way:
 \begin{equation}\label{Eq:Impl1}
    \frac{\partial \cU_\mathrm{r}}{\partial t} = -\hat{c}\,\cG \,,\,\,\,\,
      \frac{\partial}{\partial t}
      \left(
    \begin{array}{c}
    E_\mathrm{tot}\\
    \mathbf{m}_\mathrm{tot}
    \end{array} 
    \right)=0\,,
 \end{equation}
 where $E_\mathrm{tot}$ and $\mathbf{m}_\mathrm{tot}$ are
 defined in Eq. \eqref{Eq:Etotmtot}.
 We implicitly integrate the first of these
 equations while keeping $E_\mathrm{tot}$ and $\mathbf{m}_\mathrm{tot}$ constant.
 Each implicit step in the IMEX schemes
 can be written as
 \begin{equation}\label{Eq:Impl2}
     \cU_r = \cU_r'
     - s\, \Delta t^{n,q}_r\,\hat{c}\, \cG\,,
 \end{equation}
 where $s$ is a constant and $\cU_r'$
 denotes an intermediate-state value.
 Since during this step $E_\mathrm{tot}$
 and $\mathbf{m}_\mathrm{tot}$ must remain
 constant, HD fields can be defined
 as functions of the radiation fields
 and vice versa by inverting Eq.
 \eqref{Eq:Etotmtot}.
 Therefore, Equation
 \eqref{Eq:Impl2} can be solved through
 iterative methods that update either $\cU_r$ or some set of HD fields that allows the inversion
 of Eq. \eqref{Eq:Etotmtot} to obtain $\cU_r$.
 
 We implemented three
 implicit methods,
 namely Newton-Rad, Newton-HD, and fixed point (FP).
 The first two of these correspond
 to Newton methods iterating, respectively, 
 $\cU_r$ and $(p_g,\mathbf{v})^\intercal$,
 while the last one is a
 fixed-point method based
 on iterations of $\cU_r$.
 Both Newton-Rad and Newton-HD present no
 major changes with respect to their implementation
 in \cite{MelonFuksman2019}, except for the 
 different form of the Jacobian due to our
 expansion of the source terms up to order
 $\bm{\beta}^2$ (Eq. \eqref{Eq:SourceTerms}).
 Similar implementations can be found
 in  \cite{McKinney2014} and \cite{Sadowski2013}.
 The FP method was introduced in
 \cite{Palenzuela2009} in the context of resistive
 relativistic magnetohydrodynamics and implemented in \cite{MelonFuksman2019} for Rad-RMHD, having
 been firstly applied in this context in \cite{Takahashi2013}.
 This scheme is based on a linearization of Eq.
 \eqref{Eq:Impl2} achieved by writing all HD variables
 and the Eddington tensor $D^{ij}$ at a previous iteration with respect to $\cU_r$.
 In that manner, 
 $\mathcal{G}$ can be written at a given iteration $m$ as
    \begin{equation}
    \mathcal{G}^{(m)}=
    \mathcal{M}^{(m)}\cU^{(m+1)}_r+b^{(m)} ,
    \end{equation}
    where
  \begin{equation}
  \mathcal{M}=\left( \begin{array}{cc}
              \rho\kappa-\rho\chi\left(\bm\beta^2+
              \beta_k\beta_l D_{kl}\right)   
              &  \rho(\sigma-\kappa)\beta_j \\
              -\rho\sigma\beta_i -\rho\chi\beta_kD_{ik}
              & \rho \chi \delta_{ij} - 2\rho\kappa\beta_i\beta_j \\
             \end{array}    \right),
  \end{equation}
  and  $b=- \kappa\, \rho\, a_R T^4
   \left(1,\bm{\beta}\right)^\intercal$.
  Finally, $\cU_r$ can be updated as
  \begin{equation}   
    \cU^{(m+1)}_r = 
    \left( \mathcal{I} + s\, \Delta t^n \mathcal{M}^{(m)} \, \right)^{-1}
    \left( \cU'_r - s\, \Delta t^n\, b^{(m)} \right),
    \end{equation}
  after which HD fields can be updated
  by inverting Eq. \eqref{Eq:Etotmtot}
  and the process can be repeated until
  convergence is reached. A convergence
  criterion is imposed in each method by
  requiring that the relative variation of
  the iterated fields  becomes lower than a given
  threshold. To avoid accuracy issues that
  may arise when $E_r$ and $E$ are different
  by several orders of magnitude 
  \citep[see, e.g.,][]{McKinney2014},
  we have added the option of imposing
  the same criterion to the relative variations
  of $p_g$ in Newton-Rad and FP, doing the same
  with $E_r$ in Newton-HD. The results shown
  in this work have been calculated using the
  FP method, as we have verified that it is
  usually the fastest one with respect to
  the other two.

%% file: application.tex
 \section{Benchmarks and applications}\label{S:Applications}

In this section we show a series
of tests of the code's performance,
paying special attention to the 
applicability of the RSLA in
the context of protostellar disks.
All of the results shown in this
section are computed employing
HLLC solvers for both the HD and
radiation fields, using the
third-order weighted essentially non-oscillatory (WENO)
reconstruction scheme by \cite{Yamaleev2009},
and applying the IMEX1 scheme
at the radiation step.
Benchmarks where matter is
either static or does not
interact with radiation,
such as the free
streaming of beams, the
formation of shadows, 
the transport of radiative pulses under different choices of coordinates,
and the higher accuracy of our
HLLC Riemann solver for radiation
transport with
respect to the HLL solver
in Riemann problems, exhibit
no differences with the
results presented in \cite{MelonFuksman2019},
except for the fact that the
velocity of freely streaming radiation is now replaced by $\hat{c}$. Hence, we do not
show such tests in this work.
Additional performance tests
and comparisons with other works
can be found in Appendix
\ref{S:Performance}.

\subsection{Radiative shocks}\label{SS:Shocks} 

We have tested the code's
ability to reproduce 
shock waves in optically
thick media, in which
the dynamical evolution
of matter and radiation
fields is coupled. 
We have reproduced the
1D setup considered
in \cite{Ensman1994},
which is generally
used as a standard benchmark in Rad-HD
codes \citep[see, e.g.,][]{Hayes2003,Gonzalez2007,Commercon2011,Kolb2013,Colombo2019}. In this
configuration, both
matter and radiation
fields are initially
uniform in a domain given
by the interval
$[0,7\times 10^{10}]$ cm.
The initial density is 
$\rho=7.78\times10^{-10}$
g cm$^{-3}$, while the
pressure and initial
radiation fields
are set in LTE
at an initial temperature $T_1=10$ K, with $\mu=1$
and $\Gamma = 7/5$
(see Eqs. \eqref{Eq:TempIdealGas} and \eqref{Eq:EoS}). Opacities
are set in such a way that
$\kappa \rho = 3.1 \times 10^{-10}$ cm$^{-1}$, with
$\sigma=0$. A rightward-moving
shock is generated
by setting an initially negative velocity $u$,
and imposing reflective
conditions on the left
boundary.

\begin{figure}[t]
\centering
\includegraphics[width=0.49\textwidth]
{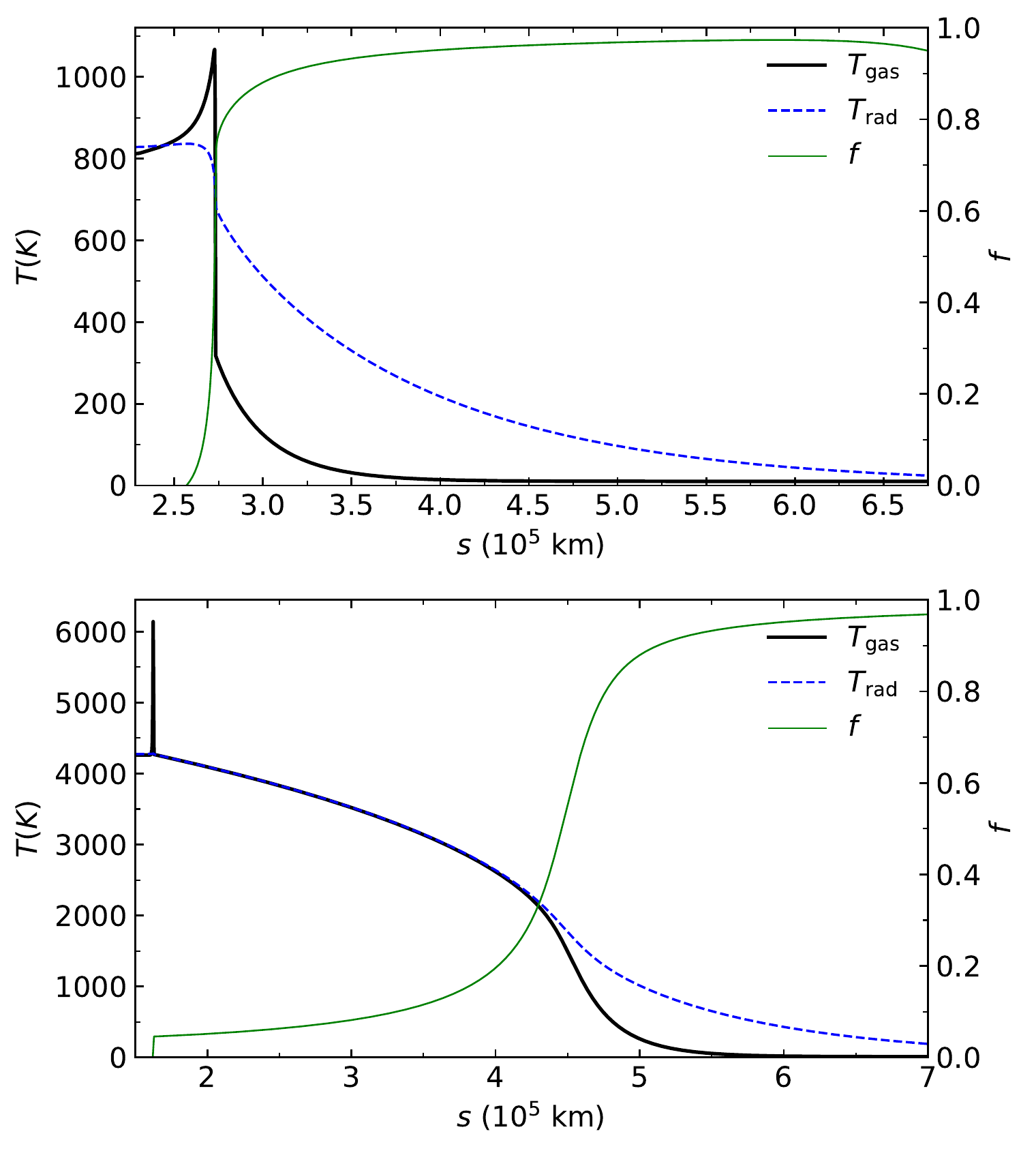}
\caption{ Gas and radiation 
temperature profiles,
here denoted by
$T_\mathrm{rad}$ and $T_\mathrm{gas}$,
for the subcritical
(top) and supercritical 
(bottom) shock problems,
shown, respectively, at
$t=3.8\times 10^4$ s
and $t=7.5\times 10^3$ s
as a function
of $s=x-ut$.
The reduced flux $f=\vert\vert\mathbf{F}_r\vert\vert/
E_r$ is also shown to illustrate 
the transition between the streaming
and the diffusion limits. The profiles have been computed with
a resolution of 2400 zones in both cases.}
\label{fig:shocks_hr}
\end{figure}

\begin{figure*}[t!]
  \centering
  \includegraphics[width=\linewidth]{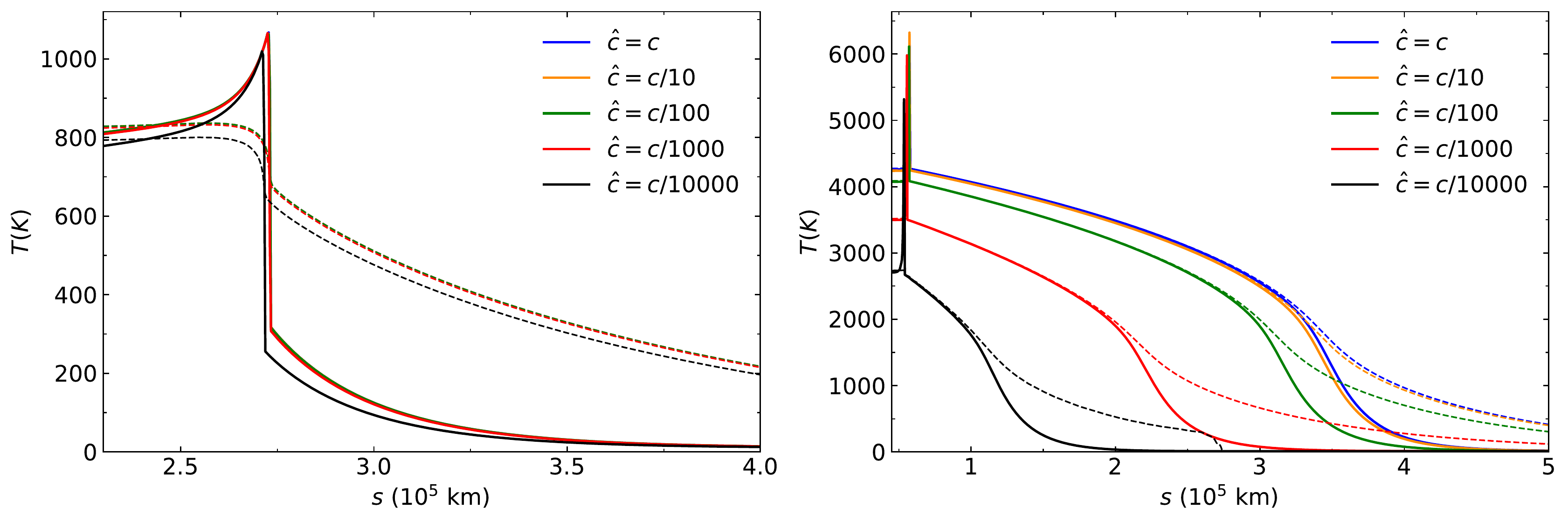}
\caption{ Gas (solid lines) and
radiation (dashed lines) temperature
profiles of the subcritical (left panel)
and supercritical (right panel) shocks
computed at a resolution of $2400$
zones for different values of $\hat{c}$. The profiles are shown at
the same times as those in Fig. \ref{fig:shocks_hr} as a function
of $s=x-ut$.}
\label{fig:shocks_cr}
\end{figure*}

Radiative shocks are
extensively studied, e.g.,
in \cite{Zeldovich1967} and \cite{Mihalas}. In the general case,
radiation escaping from the
shock front will cause
the pre-shock region to
raise its temperature
until reaching a value $T_-$ at the shock front.
In the shocked region,
the temperature decreases
from its maximum value 
$T_+$ at the shock front
until its 
post-shock value $T_2$.
The behavior of the
solutions depend of the
fluid's velocity,
in our case parameterized
by $u$.
For values of $\vert u\vert$ below
a critical value $ u_\mathrm{cr}$,
the resulting
temperature
profile verifies
$T_-<T_+$, and
the produced shock
is said to be
\emph{subcritical}. 
For higher velocities
shocks are said to
be \emph{critical} if
$\vert u\vert= u_\mathrm{cr}$
and \emph{supercritical}
if $\vert u\vert> u_\mathrm{cr}$,
and always verify $T_-=T_+$. % Maybe add temperature estimates

We employed two values
of $u$ given by $-6$ and
$-20$ km s$^{-1}$, which
correspond respectively
to subcritical and
supercritical velocities.
We produced numerical
solutions starting from
both conditions using
in every case
a uniform grid of $2048$
zones, setting $\hat{c}=c$
to avoid inaccuracies produced
by the RSLA. These results
are shown in Fig. \ref{fig:shocks_hr},
in which we show the obtained
temperature profiles at 
$t=3.8\times 10^4$ s
and $t=7.5\times 10^3$ s
for the subcritical and
supercritical shock, respectively.
We have as well computed the
radiation temperature $T_\mathrm{rad}$,
defined as $T_\mathrm{rad}=
\left(E_r/a_R
\right)^{1/4}$, which corresponds
to the equilibrium temperature
in LTE. In the same figure
we have represented the reduced
flux
$f=\vert\vert\mathbf{F}_r\vert\vert/
E_r$.
All profiles have been plotted
as a function of $s=x-ut$
for comparison with the
mentioned works.
The structure of the
temperature in the
precursor, namely the
heated pre-shocked region,
differs in both cases. 
In the subcritical shock there is
an abrupt transition from the
diffusion to the streaming limit.
In the entire precursor, $f$
remains above $0.75$, and
$T_\mathrm{rad}$ exceeds the
gas temperature. This transition
is much smoother in the supercritical shock, where
$T_\mathrm{rad}=T$ and $f\leq 0.3$
in a large portion of the precursor.
We obtain $T_+=1067$ K,
$T_-=317$ K, and $T_2=812$ K
for the subcritical shock,
and $T_+=6140$ K and 
$T_2=4260$ K in the supercritical
shock. In the first of these
cases, all temperatures
except $T_2$ exceed
those obtained with FLD at the
same resolution \citep[see, e.g.,][]{Commercon2011},
and the same holds for $T_+$
in the second case. Differences
can also be seen in the precursors,
which have a generally larger
spatial extent with the M1
closure than with FLD
\citep[see also][]{Gonzalez2007}.
It is not possible from
this comparison to conclude that one of the two methods yields more
accurate results in
this particular case,
as both of them rely on
an approximate closure.
In general, both methods
produce similar results
in 1D, whereas the M1
method outperforms FLD
in multidimensional anisotropic setups,
e.g., involving beams
or shadows. A better 
comparison in this 1D
case would require
the employment of
radiative transfer techniques that do not depend on the choice of
a closure prescription
\citep[see, e.g.][]{Davis2012}, which
is beyond the scope
of this paper.

We have used this test to study the
limits of the RSLA formalism when applied to nonequilibrium systems.
To do this, we have performed
the same tests
using different values of 
$\hat{c}$
of the form $c/10^n$, with
$n\in [0,4]$.
The resulting $T$ and 
$T_\mathrm{rad}$ profiles
are shown in Fig. \ref{fig:shocks_cr}
at the same times as those in Fig. \ref{fig:shocks_hr}.
The obtained temperatures are
systematically smaller than their
values with $\hat{c}=c$ as
$\hat{c}$ is reduced
as a result of
the nonconservation of
the total energy in
the RSLA. Using the first of
Eqs. \eqref{Eq:Etotmtot}
together with the condition
$\Delta E_\mathrm{tot}=0$
verified in the implicit step
(Eq. \eqref{Eq:Impl1}),
we can write the variation
of the total energy
as
\begin{equation}\label{Eq:EnergyLoss}
    \Delta(E+E_r)=
(1-c/\hat{c})\Delta E_r\,,
\end{equation}
which is negative unless
$\hat{c}=c$, since 
$\Delta E_r>0$ in this case. Therefore,
more energy will be artificially
lost for smaller values of $\hat{c}$.
Since the conversion of
kinetic energy into thermal energy
is faster in the supercritical
shock than in the subcritical
shock, this effect is more 
important in the former, while
the latter
can be reproduced by the RSLA
for smaller $\hat{c}$ values.
As an example, the relative L$_1$-norm
difference between the obtained
$T$ with $\hat{c}=c$ and
$\hat{c}/1000$ is of $0.2\%$
in the subcritical shock and
$41\%$ in the other case.

We can give rough estimates for 
the range of values of $\hat{c}$
in which the RSLA is applicable
by applying Eq. \eqref{Eq:ConstraintC} 
computing $\tau_\mathrm{max}$ as
the total optical depth of the
domain and replacing $v_\mathrm{max}$ by the maximum value of $\vert v_x\vert+c_s$, where $c_s$ is the
fluid's sound speed.
Using the profiles obtained with
$\hat{c}=c$, this yields the
conditions
$\hat{c}\gg c/2068$ and
$\hat{c}\gg c/543$ for the
subcritical and supercritical
shocks respectively.
This criterion
alone does not explain
why in the subcritical case the solutions depart more
than $1\%$ from the
$c=1$ solution only
for $\hat{c}$ below
its approximate limiting value, whereas in the supercritical case they do so
for $\hat{c}=c/100$,
which is still about five times larger than $c/543$. 
However, this timescale analysis does not 
contemplate the error
introduced by the RSLA
when gas energy is continuously injected into the system from the boundaries and converted into radiation energy,
which can cause a significant energy loss 
for sufficiently low $\hat{c}$ (see Eq. \eqref{Eq:EnergyLoss}). 
To obtain an approximate condition for the validity of the RSLA
in this case, we estimate the ratio of lost energy to total kinetic energy converted into internal energy at the left boundary as
$\max\left( \Delta(E+E_r)/(\rho \epsilon+E_r)
\right)$,
where we use Eq. 
\eqref{Eq:EnergyLoss}
to compute $\Delta(E+E_r)$
taking $\Delta E_r \approx E_r$. Requiring this
ratio to be much smaller
than $1$ and
approximating $1-c/\hat{c}\approx -c/\hat{c}$,
this gives the condition
$\hat{c}\gg c/50000$ for the subcritical shock and
$\hat{c}\gg c/400$ for the supercritical shock.
Therefore, errors
above $1\%$ can be seen 
in both shocks when
$\hat{c}$ is about $4-5$
times larger than these limiting values. However,
as is the case for Eq.
\eqref{Eq:ConstraintC},
these are approximate
relations, and
optimal values of
$\hat{c}$ are better determined in general through testing.

\subsection{Diffusion in disk atmospheres}\label{SS:1Ddiff} 

As a first application of the code in
the context of protoplanetary disks,
we have considered a one-dimensional
setup representing a vertical slice
of a disk at a radius $R=5$ AU with
respect to a central star of mass $M_\odot$.
We have used this setup to test the
effect of the RSLA on the timescales
corresponding to processes of viscous
heating and radiative diffusion.
Similar tests have been performed e.g. in \cite{Zhu2020}.

\begin{figure*}[t!]
\centering
  \includegraphics[width=\linewidth]{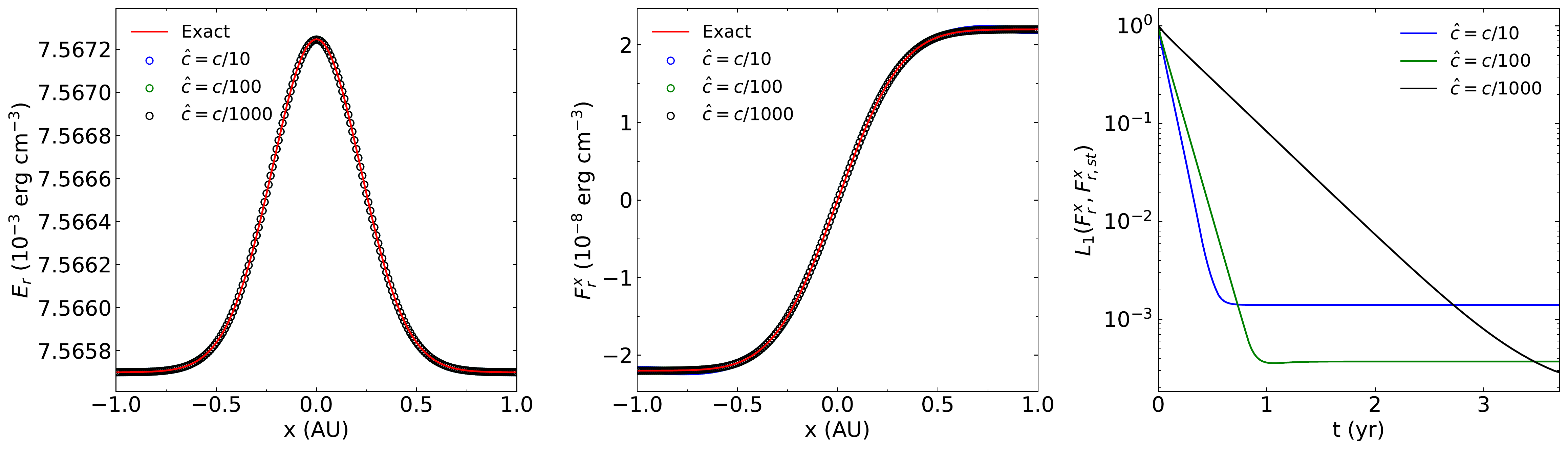}
\caption{Left: radiation energy density
at $t=3.8$ yr in the diffusion test
for different values of $\hat{c}$,
compared with the exact semi-analytical stationary solution.
Middle: same as the plot in the left panel, this time showing the equilibrium radiative flux.
Right: relative L$_1$-norm difference
between the numerical and analytical values of $F^x_r$ as a function of time. }
\label{fig:diffusion}
\end{figure*}

We define this problem in a domain
given by the interval $[-1,1]$ AU,
where we set a Gaussian
density profile defined as
\begin{equation}
    \rho(x) = \rho_0
    \exp{(-x^2/2H^2)}
    +\rho_\mathrm{min}\,,
\end{equation}
where $\rho_0=10^{-10}$ g cm$^{-3}$
and $\rho_\mathrm{min}=10^{-10}$ $\rho_0$, while the pressure scale
height $H$ is defined in such a way that $H/R=0.05$. Such a distribution represents
a vertical density profile
resulting from the balance between the gravitational force of the star and the
internal pressure of the disk.
Since in this case we are solely interested
in the diffusion of radiative energy, 
we neglect gravity and all 
advection terms for energy-momentum
and matter. The resulting
evolution equations are therefore
\begin{equation} \label{Eq:DiffTest}
\begin{split}
    \frac{\partial E}{\partial t}
    &= c G^0 + S_E \\
    \frac{1}{\hat{c}}\frac{\partial E_r}{\partial t}+
    \frac{\partial F^x_r}{\partial x}
    &= -G^0 \\
\frac{1}{\hat{c}}\frac{\partial F^x_r}{\partial t}+
\frac{\partial P^{xx}_r}{\partial x}
&= -G^x\,.
\end{split}
\end{equation}
Following the $\alpha$ prescription
by \cite{ShakuraSunyaev},
we compute the viscous heating term
as
$S_E=\frac{9}{4}\alpha\Omega_K c_s^2 \rho$, where $\alpha=10^{-3}$,
$\Omega_K$ is the Keplerian angular
velocity at $5$ AU,
and $c_s$ is the speed of sound
computed at the initial uniform
temperature $T_0=1000$ K. We 
set the absorption opacity
$\kappa=0.1$ cm$^2$ g$^{-1}$,
zero scattering, $\mu=2.35$,
and an adiabatic
index $\Gamma = 1.41$, corresponding
to typical values for solar composition
\citep{DeCampli1978}. Initial
LTE conditions with $T=T_0$ are imposed
in the entire domain at $t=0$
and at the boundaries for $t>0$,
while zero-gradient boundary
conditions are imposed on
$F^x_r$. 

The final state of this system corresponds to a stationary configuration in which viscous heating and radiation diffusion are in equilibrium. This solution can be obtained semi-analytically by
setting all time derivatives in Eq.
\eqref{Eq:DiffTest} to $0$,
which leads to $G^0=-S_E/c$.
Since $S_E$ is a known function of $x$,
the second of these equations can be numerically integrated to yield $F^x_r$, for which we
use the condition $\partial_x F^x_r=0$
at $x=0$. The third equation can
be in turn integrated to yield
$P^{xx}_r(x)$ using the values of
$E_r=a_R T_0^4$ and $F^x_r$ at
one of the domain boundaries.
Lastly, the values of
$P^{xx}_r(x)(E_r,F^x_r)$ can be
inverted to obtain $E_r$. This
inversion leads to unique $E_r$
solutions provided $F^x_r/E_r< 3/7$
\citep[][]{MelonFuksman2019},
which is satisfied since in our case
$F^x_r/E_r\sim 10^{-5}$.

Simulations have been run
taking $\hat{c}=c/10$, $c/100$,
and $c/1000$,
at a resolution of
$201$ zones in each case.
The resulting $E_r$ and $F^x_r$
profiles are shown in Fig.
\ref{fig:diffusion} at $t=3.8$ yr
$= 1.06\, \Omega_K^{-1}$,
together with the described
semi-analytical solution,
where a good agreement is
obtained in each case.

In the right panel
of Fig. \ref{fig:diffusion}
we have plotted as a function of time
the L$_1$-norm relative difference between the numerical values of $F^x_r$ and the
stationary semi-analytical solution.
We can see in that figure that
the stationary solution is reached
at earlier times for larger  $\hat{c}$,
and that smaller $\hat{c}$ values lead
to more accurate stationary
solutions. The reason for this is
that a slower evolution of the
system leads to smaller values of
the time derivatives, which reduces
the imbalance between
$cG^0$ and $S_E$ caused by
operator splitting error.
We estimated the
timescale in which the radiative flux
reaches its final configuration by computing the 
initial slopes of these curves, obtaining $t_\mathrm{eq}=0.045$ $\Omega_K^{-1}$, 
$0.062$ $\Omega_K^{-1}$,
and $0.224$ $\Omega_K^{-1}$ for
$\hat{c}=c/10$, $c/100$, and $c/1000$, respectively.

\subsection{Convective instability in protoplanetary disks}\label{SS:Convection} 

We now turn to a scenario in which convective
vertical flows are spontaneously produced in a protoplanetary disk. Convection occurs when
vertical superadiabatic temperature gradients
are created, which in our case happens as a product of the balance between viscous heating, adiabatic compression, and
radiative diffusion in the disk.
Such unstable temperature gradients
are difficult to sustain in
time, as reviewed in \cite{Klahr2007}, and it is unknown whether
they can be maintained
through some support
mechanism such as the formation of strong spiral
shocks caused by orbiting planets
\citep[][]{Lyra2016}. However, convective energy transport might still regulate the formation of vertically adiabatic
stratifications,
which aids the growth of
other turbulence-driving mechanisms such as the vertical shear instability
\citep[see][]{Pfeil2019}. Therefore,
vertical convection might
still have a role
in the development of turbulence
and angular momentum transport
in the dead zones of protoplanetary
disks, where the low ionization
degrees render the magnetorotational
instability inefficient \citep{Gammie1996}.

\begin{figure*}[t!]
\centering 
  \centering
  \includegraphics[width=.49\linewidth]{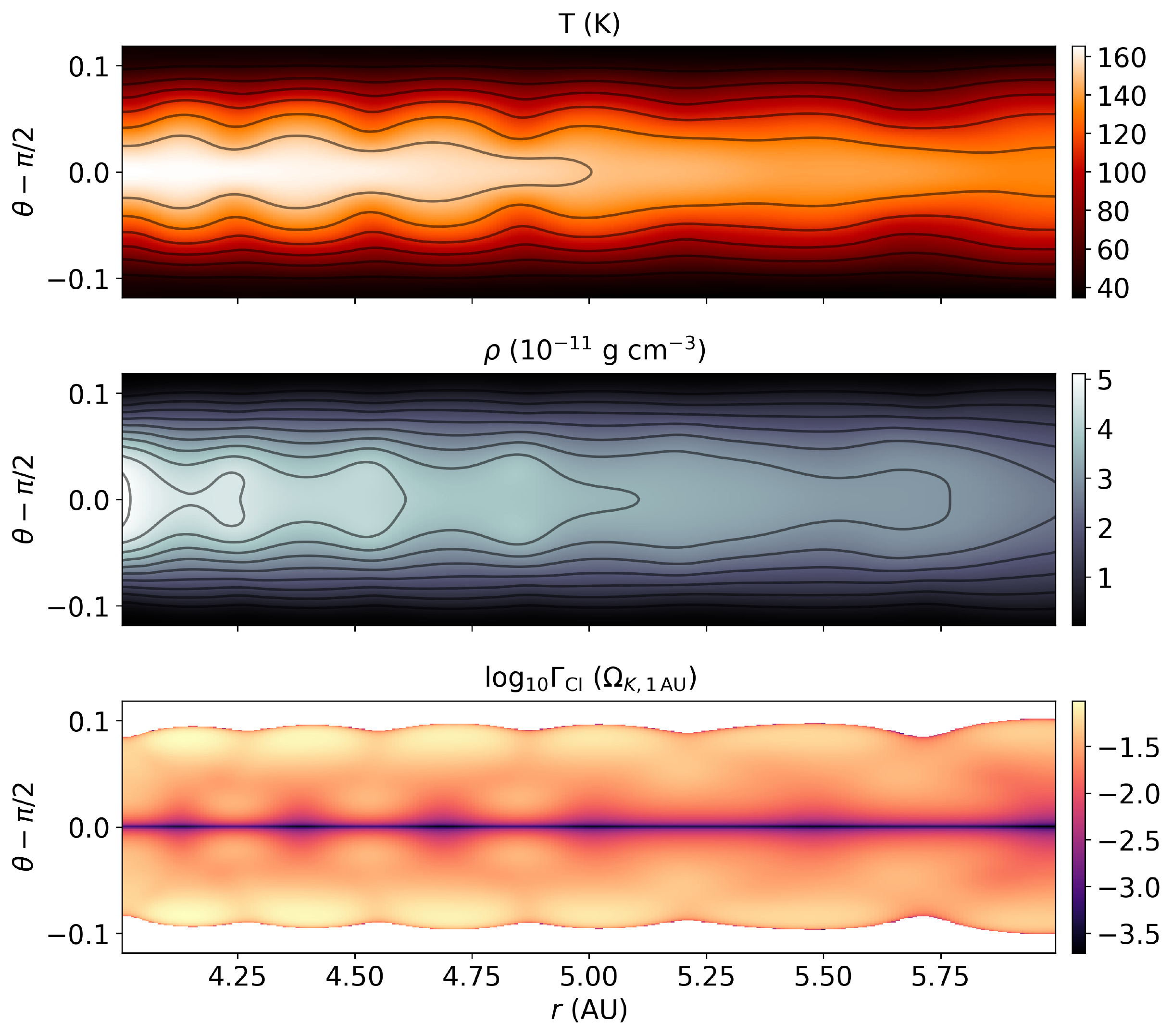}
  \centering
  \includegraphics[width=.49\linewidth]{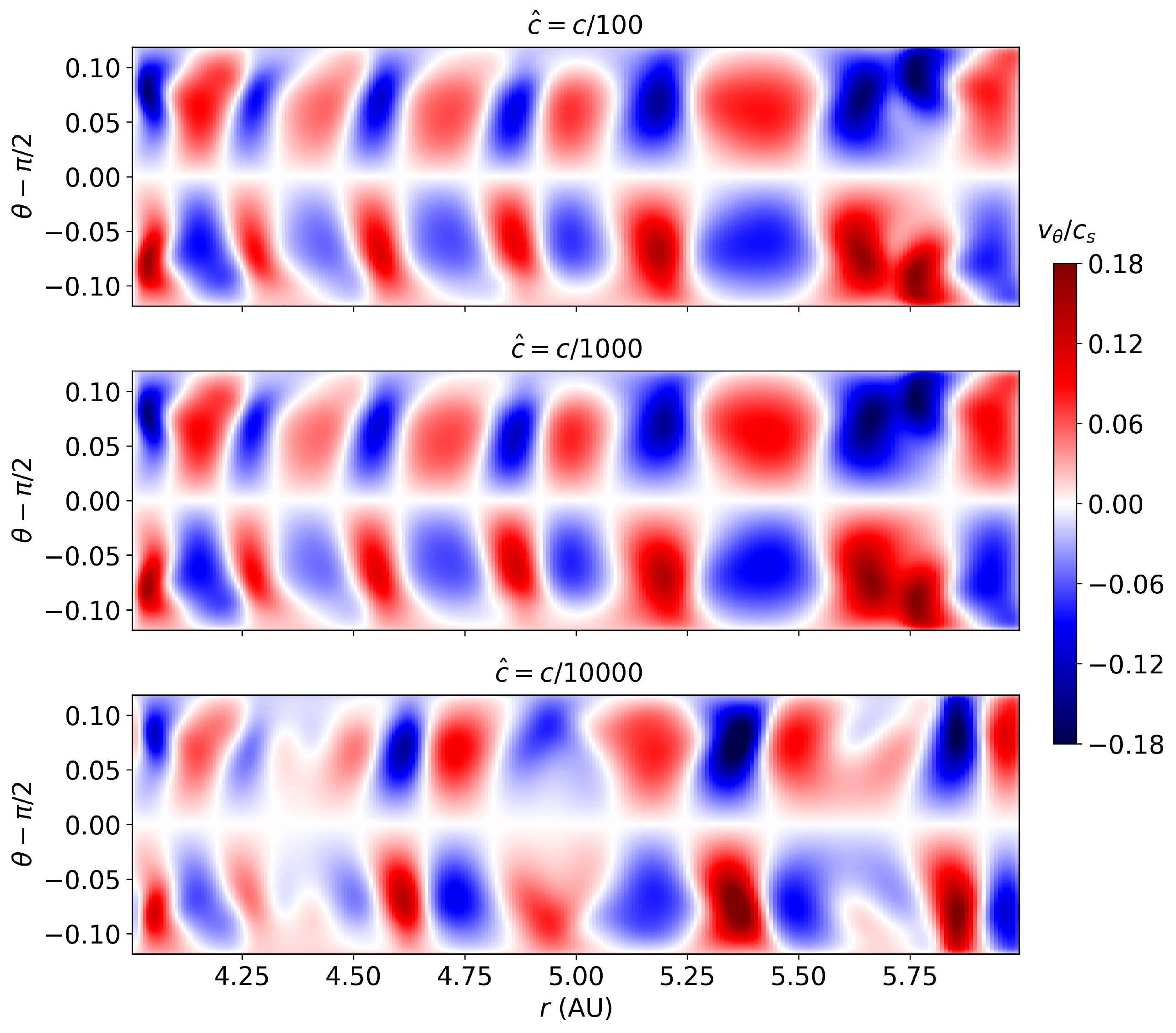}
\caption{ Left: temperature (upper panel),
density (middle panel), and logarithm
of the growth rate of the convective
instability (lower panel) as a
function of $(r,\theta)$ in
run C3, plotted at $t=200$ $T_0$. Contour lines
are included in the temperature profile
every $20$ K between $50$ and $160$ K,
as well as in the density profile 
every $0.5$ g cm$^{-3}$ between $0$ and $5$
g cm$^{-3}$.
Right: values of $v_\theta/c_s$ after
200 orbits in runs C2, C3, and C4.}
\label{fig:convcells}
\end{figure*}

We applied our code to describe a
convectively unstable setup, with a
particular focus on how the RSLA affects
the evolution of the instability. 
We consider the case of
an axisymmetric disk, and
solve the Rad-HD equations in a
2D grid using spherical
coordinates $(r,\theta)$.
Similar configurations have been considered
in \cite{Cabot1996} and \cite{Klahr1999}.
This time we solve the full Rad-HD equations,
including the viscous heating source terms
given by Eq. \eqref{Eq:ViscSourceTerms}
and the gravitational potential of a solar
mass star, given by $\Phi(r)=-G M_\odot/r$.
We set an initial vertically isothermal configuration at LTE, with the
density and rotational
angular velocity given by
\begin{equation}\label{Eq:RhoOmegaDisk}
\begin{split}
    \rho(R,z)&=\rho_0 \left(
                    \frac{R}{R_0}
                        \right)^p
    \exp \left(\frac{R^2}{H^2}
         \left[
         \frac{R}{\sqrt{R^2+z^2}}-1
         \right]
    \right) \\
    \Omega(R,z)&=\Omega_K
    \left[
    (1+q)-\frac{q\,R}{\sqrt{R^2+z^2}}
    +(p+q)\frac{H^2}{R^2}
    \right]\,,
\end{split}
\end{equation}
\citep[see, e.g.,][]{Fromang2011},
where $(R,z)=(r\sin \theta,r \cos \theta)$
are the cylindrical radius and height, while
$\rho_0=10^{-9}$ g cm$^{-3}$, $R_0=1$ AU,
$p=-2$, $q=-1/2$, and
$\Omega_K=\sqrt{G M_\odot/R^3}$ is the midplane
Keplerian angular velocity.
The pressure scale height is computed
as $H=H_0(R/R_0)^{(q+3)/2}$, where
$H_0/R_0=0.035$. With the chosen value of
$q$, this gives an increasing $H/R$ ratio 
proportional to $R^{1/4}$.
The gas pressure is computed
as $p_g=\rho c_s^2$, where $c_s$ is the local
sound speed, estimated as $c_s=H \Omega_K$.
In this way, the initial temperature decreases radially as $T\propto R^{q}$. 

Accretion disks are unstable to thermal
vertical convection under the condition
that entropy decreases away from the disk
midplane, i.e.,
\begin{equation}
\frac{\partial S}{\partial \vert z\vert} = 
C_v \frac{\partial}{\partial \vert z \vert}\log \left( \frac{p_g}{\rho^\Gamma}
\right) <0\,,
\end{equation}
where $S$ is the specific entropy and 
$C_v$ is the specific heat at constant volume.
\cite{LinPapaloizou1980} have shown that such a gradient can be obtained in a disk that radiates vertically while decreasing its internal energy and consequently shrinking.
Considering an absorption opacity of the
form $\kappa = \kappa_0 T^\beta$, they have
derived the criterion
\begin{equation}\label{Eq:CondLinPapaloizou}
    \frac{1}{4-\beta} \geq
    \frac{\Gamma-1}{\Gamma}
\end{equation}
for the disk to be vertically unstable to
convection. For our model, we have used
the absorption opacity law by \cite{BellLin1994},
which consists of a series of broken
power laws of the form
$\kappa=\kappa_0\, \rho^\alpha T^\beta$
corresponding to the absorption of
millimeter-sized grains in different
temperature regimes.
For temperatures of at most
a few hundred Kelvin, the absorption opacity is
dominated by ice grains if $T\lesssim 160$ K,
in which case $\kappa_0 = 2\times 10^{-4}$ cm$^2$ g$^{-1}$, $\alpha=0$, and $\beta = 2$,
while for higher temperatures metal grains
dominate the absorption, and the
parameters are
$\kappa=0.1$ cm$^2$ g$^{-1}$, 
$\alpha=0$, and $\beta = 1/2$.
For $\Gamma= 1.41$,
we obtain that condition
\eqref{Eq:CondLinPapaloizou} is only
satisfied below the ice line. For this
reason, we have chosen our parameters
in such a way that the temperatures do
not overpass this threshold, but remain high enough that a superadiabatic temperature
gradient is produced before all energy
is radiated away. We also set zero scattering
and $\mu=2.35$.

In order to satisfy these conditions,
we model the disk in the region
$(r,\theta)\in [4,6]\,\mathrm{AU}\times[\pi/2-0.12,\pi/2+0.12]$, with a viscosity determined by the $\alpha$ prescription
\citep{ShakuraSunyaev}
as $\nu = \alpha\, \Omega_K^{-1} p_g/\rho $
(see Eq. \eqref{Eq:ViscTensor}). 
We impose zero-gradient conditions for $p_g$
in the inner and outer radial boundaries,
setting $v_\phi=\Omega R$ as in Eq. \eqref{Eq:RhoOmegaDisk}
and $v_r=v_\theta=0$, in such a way that
the mass flow through these boundaries is zero.
In the poloidal direction reflective conditions
are applied on all HD fields.
We set the radiation flux to zero gradient
except in the case of radiation inflow, in
which case impose reflective conditions.
The radiation energy is set to zero gradient
in the radial direction, whereas in the
poloidal direction we fix it to
$E_r=a_r T^4_\mathrm{min}$ with
$T_\mathrm{min}=10$ K $\ll T$ in all
ghost cells. This is essential to ensure
that the radiated energy leaves the system
instead of accumulating in the domain,
eventually leading to the thermalization
of the system.

\begin{figure}[t]
  \centering
  \includegraphics[width=\linewidth]{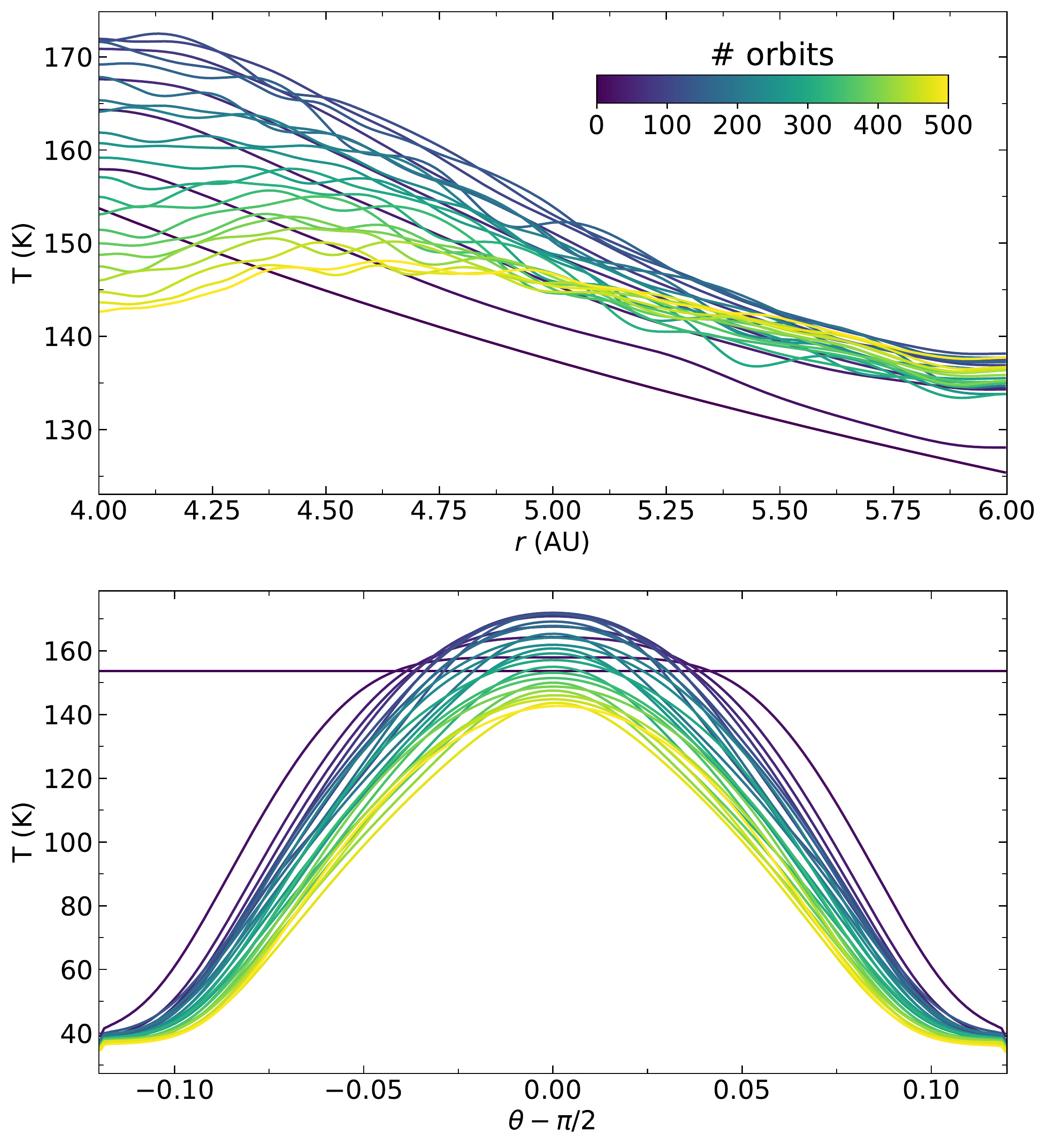}
\caption{Temperature profiles in run C3
at $\theta=\pi/2$ (top panel) and
$r=4$ AU (bottom panel),
plotted every $20$ orbits. 
The color scale indicates the current
number of orbits for each profile.
}
\label{fig:convT}
\end{figure}

\begin{figure*}[t!]
\centering
\includegraphics[width=\linewidth]{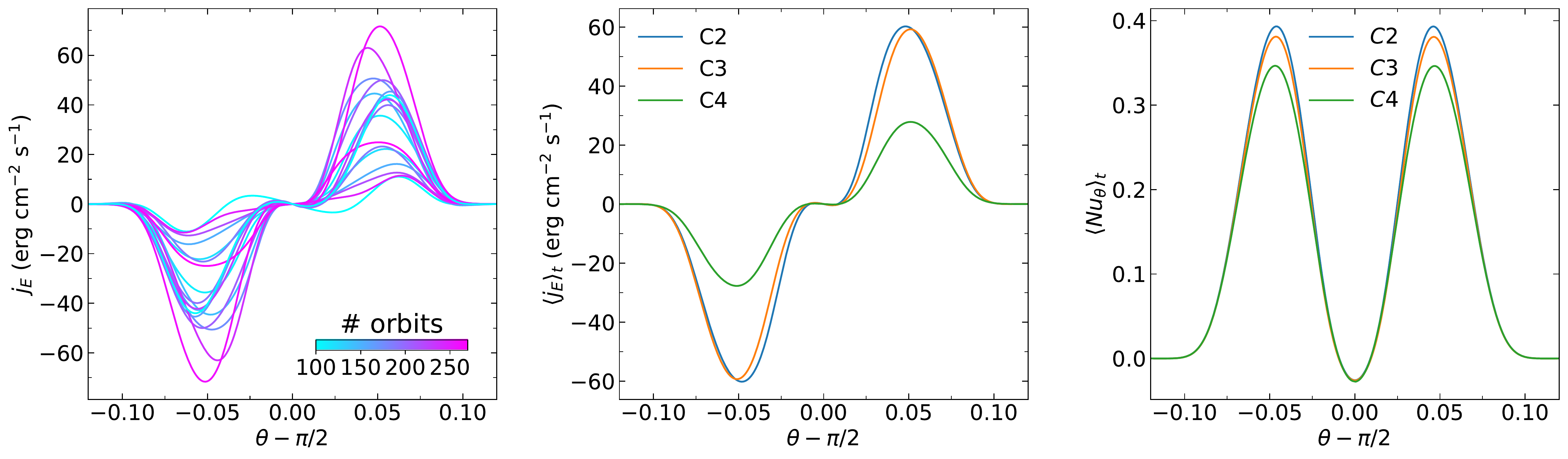}
\caption{ Left:
vertical convective heat flux $j_E$
as a function of time, plotted
every $10$ orbits between
$100$ and $270$ orbits. Center:
Time-averaged $j_E$ profiles for
runs C2, C3, and C4. Right: time-averaged
Nusselt number for the same runs.
}
\label{fig:convflux}
\end{figure*}

Computations have been run on a grid
of $256^2$ zones logarithmically spaced in the radial direction,
using $C_a=0.3$ for both the radiation
and HD fields. 
We have performed three different runs of
this test using in each case a different
value of $\hat{c}$, namely $\hat{c}=c/100$,
$c/1000$, and $c/10000$. We refer to these 
simulations as C2, C3, and C4, respectively.
We ran C2 and C4 for a total of 275
orbits and C3 for 500 orbits, where we
define an orbit as the Keplerian 
period at $1$ AU, i.e., $T_0=2\pi/\Omega_{K,1\,\mathrm{AU}}$.

In every run, the system goes through an
initial relaxation phase lasting a few tenths
of orbits, in which radially oriented sound
waves can be observed in the velocity profiles.
The entropy gradient becomes unstable close
to both vertical boundaries from the first
orbit. The unstable regions migrate toward
the midplane until merging at $t\approx 70\, T_0$.
At this point, vertical convective cells
can begin to be observed in the velocity
profile, and at $t\approx 100\, T_0$ they
become evident in the density and temperature
profiles as well. This can be seen in Fig. 
\ref{fig:convcells}, where we have plotted the
temperature and density profiles in run C3
at $t=200\, T_0$. In the same figure
we have plotted $v_\theta/c_s$, i.e.,
the projection of the velocity onto $\hat{\mathbf{e}}_\theta$
normalized by the local sound
speed. In this case the profile evidences a series of radially distributed expansive and
compressive zones.
The temperature profile has a larger
scale height in the expansive zones,
and vice versa, whereas the density
scale height is larger in the compressive ones. Convection
cells continuously migrate
in the radial direction,
interacting with each other
and sometimes merging.

We can see in the $v_\theta/c_s$
profiles that convective cells occupy
almost the entire domain, with a vertical
size limited by the size of the domain, and
a typical radial extension of about a pressure scale height, here roughly $0.1$ to $0.3$ AU. In C2 and C3
the average maximum $v_\theta/c_s$ is
$0.26$, whereas in C4 this value is
reduced to $0.19$. We compare the velocity profiles at 200 orbits in Fig.
\ref{fig:convcells}, where it
can be seen that the profiles in
C2 and C3 are almost identical,
while differences can be observed
with respect to C4. 

The computed vertical velocities can be used
to verify the constraint on the value of
$\hat{c}$ given by Eq. \eqref{Eq:ConstraintC}.
Using the maximum values of $v_\theta$ in C2
as $v_\mathrm{max}$ and computing the vertical
optical depth from the disk midplane,
we obtain the constraint $\hat{c}\gg c/8000$, which is not satisfied by C4. In C3, on the
other hand, the value of $\hat{c}$
exceeds the limit value by a factor 8.

The unstable region of the domain is shown
in Fig. \ref{fig:convcells} at 200 orbits
in C3. In the same figure we have indicated
the growth rate of the instability
at each position, calculated in terms
of the vertical
Brunt-V\"ais\"al\"a frequency $N_z$
\citep[see, e.g.,][]{Rudiger2002} as
\begin{equation}
\Gamma_\mathrm{CI}=
\sqrt{-N_z^2}=
\sqrt{\frac{1}{\Gamma\rho}
\frac{\partial p_g}{\partial z}
\frac{\partial}{\partial z}
\log\left(
\frac{p_g}{\rho^\Gamma}
\right)}\,.
\end{equation}
We see that the unstable region occupies
the entire radial extension of the domain
and almost its entire angular extension.
The growth rate increases for larger
heights at each $r$, reaching at that
time a maximum value of
of $9.7\times 10^{-2}$ $T_0^{-1}$. 

In Figure \ref{fig:convT} we show two
series of 1D temperature profiles,
one of them at $r=4$ AU and the other one
at $\theta=\pi/2$, computed in C3 every
20 orbits. The disk midplane goes through
an initial heating phase that lasts
approximately 100 orbits, reaching
a maximum temperature of $\sim 170$ K
at $r=4$ AU. During this phase, the
outer boundaries of the disk begin
to radiate out internal energy,
steepening the vertical temperature gradient
until it becomes unstable and triggers
the convective motion. Approximately at that
time, the midplane temperature profile
begins to flatten as the internal energy
of the higher-temperature regions escapes
the system through radiative diffusion.
At $t=500$ $T_0$, convection is still
occurring and the disk is steadily
cooling down while the unstable region
slowly begins to shrink.

Throughout the disk evolution, momentum and
entropy are vertically transported  through
convection. To measure the vertical entropy
transport, we define the convective
heat flux at a given $\theta$ as
\begin{equation}
    j_E(t,\theta)=
    \langle \epsilon' (\rho v_\theta)' \rangle_r\,,
\end{equation}
where $\langle\cdot\rangle_r$ represents average
in $r$, and primed quantities correspond
to deviations with respect to the average,
i.e.,
\begin{equation}
    v'(t,r,\theta) = v(t,r,\theta) - \langle v \rangle_r(t,\theta)\,,
\end{equation}
where $v$ is any given field. The behavior
of $j_E$ as a function of time
is oscillatory, as can be seen in Fig. 
\ref{fig:convflux}, where we have plotted
$j_E$ in run C3 as a function of $\theta$
every 10 orbits from $100$ to $270$ orbits.
It can already be seen in this figure that
transport occurs predominantly outwards.
This can be quantified in a more precise
way by computing the time average of
$j_E$, which we denote as
$\langle j_E \rangle_t$. In Fig. \ref{fig:convflux} we show
these averages between 100 and 270
orbits for all runs. We obtain similar
functions for C2 and C3, whereas in C4
the maximum flux is reduced to
approximately to $50\%$ of its
value in C2 and C3.

The above results show that convective
energy transport becomes more inefficient
when the speed of light is reduced.
Naturally, the same happens with the
radiative energy transport. To compare
the effect of the reduction of $\hat{c}$
on both mechanisms, we have quantified
the ratio between convective and
radiative energy transport analog to \cite{Bell1997} in the spirit of a Nusselt number, defined in this case as
\begin{equation}
\mathrm{Nu}_\theta(\theta,t) =
\frac{j_E}{\langle F^\theta_r \rangle_r}\,.
\end{equation}
Note that the classical Nusselt number gives the enhancement factor 
of total heat transport if convection adds to conduction, which can never be smaller than 1. As we do not determine the heat transport for the radiation transport only case, we slightly modified our definition of $\mathrm{Nu}_\theta$ as ratio of conductive transport over radiation transport, while both are active, and thus our $\mathrm{Nu}_\theta$ can obtain values of less than one.
We computed the time-averaged
value of $\mathrm{Nu}_\theta$ for all runs, shown in Fig.
\ref{fig:convflux}. We observe differences
in $\langle \mathrm{Nu}_\theta \rangle_t$ close to its maximum
value, which tends to decrease for increasing $\hat{c}$. We obtain $\max\langle \mathrm{Nu}_\theta \rangle_t=0.39$ in C2, $0.38$ in C3, and 
$0.34$ in C4. We conclude that
the RSLA can reproduce the main
features of this model for
$\hat{c}\geq c/1000$.

\begin{figure*}[t!]
\centering
  \includegraphics[width=\linewidth]{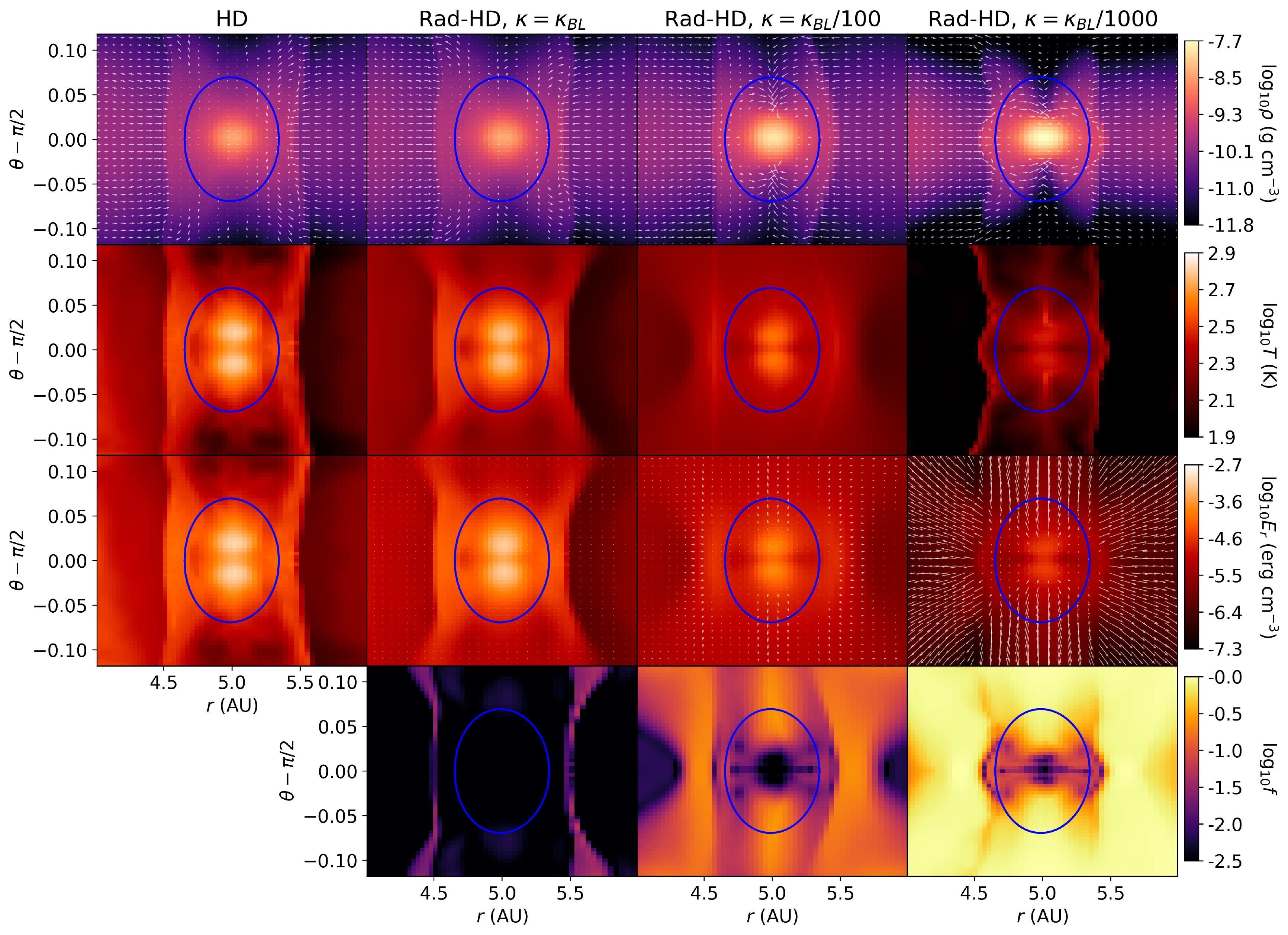}
\caption{ 
Vertical slices at the planet's location in the planet-disk
interaction test, shown at $5.5$ orbits.
From top to bottom, the logarithms of mass density, gas
temperature, radiation energy density, and reduced radiative
flux are represented in color scale. From left to right,
the results are shown for runs DP\_HD, DP\_K1, DP\_K100, and
DP\_K1000. The blue curve indicates in each case the location of the planet's Hill sphere. White arrows representing
the velocity field are superimposed in the density plots,
where we have used the same scale for all runs. In the same
way, the poloidal components of
$\mathbf{f}=\mathbf{F}_r/E_r$ are represented in the $E_r$ plots
for DP\_K1, DP\_K100, and DP\_K1000,
using in each case the same scale.
}
\label{fig:planetedgeon}
\end{figure*}

\begin{figure*}[t!]
\centering
  \includegraphics[width=\linewidth]{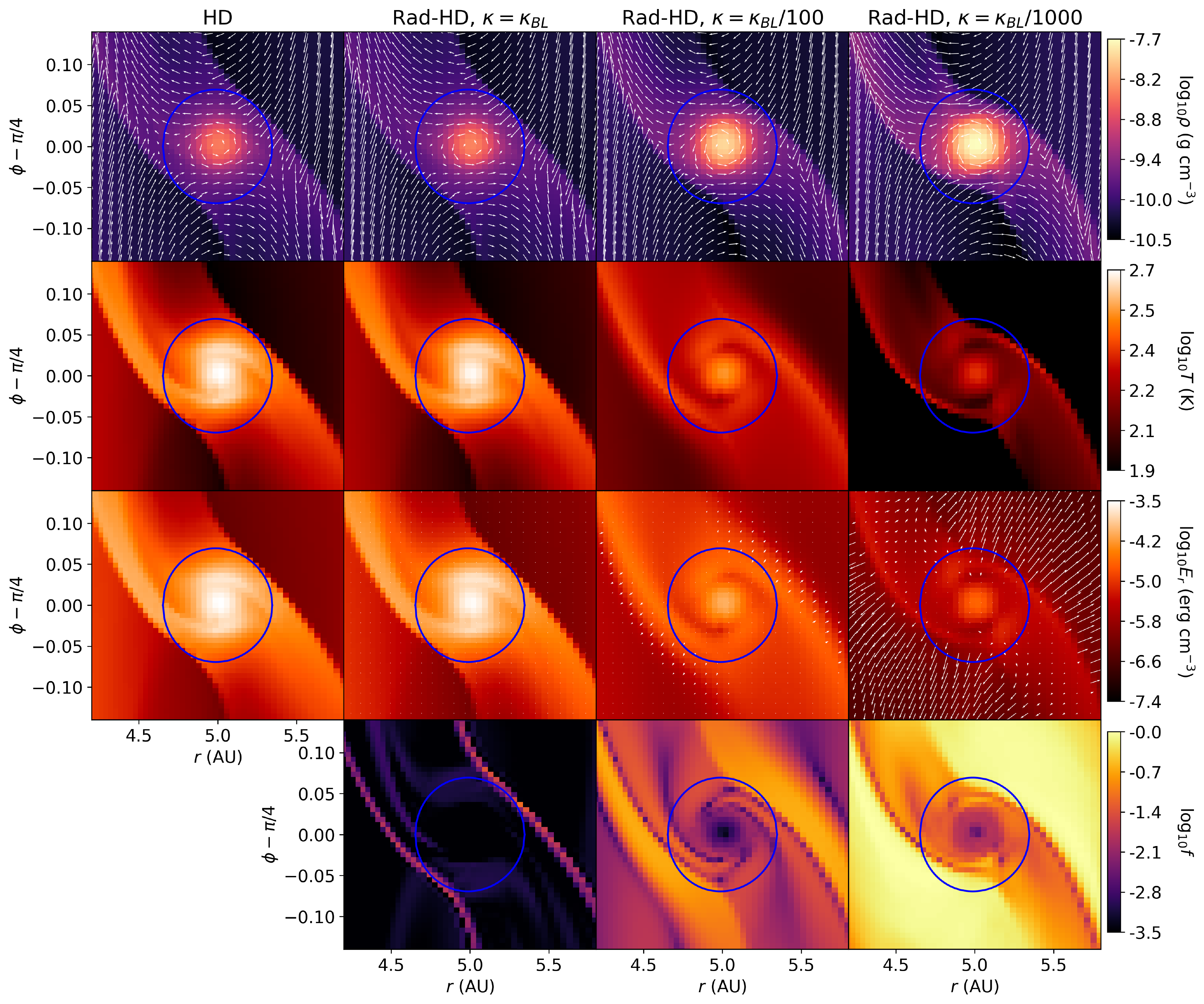}
\caption{ Same as Fig. \ref{fig:planetedgeon}, this time
showing horizontal slices of the represented fields at the
planet's location.
}
\label{fig:planettop}
\end{figure*}

\begin{figure*}[t]
\centering
\includegraphics[width=0.9\textwidth]
{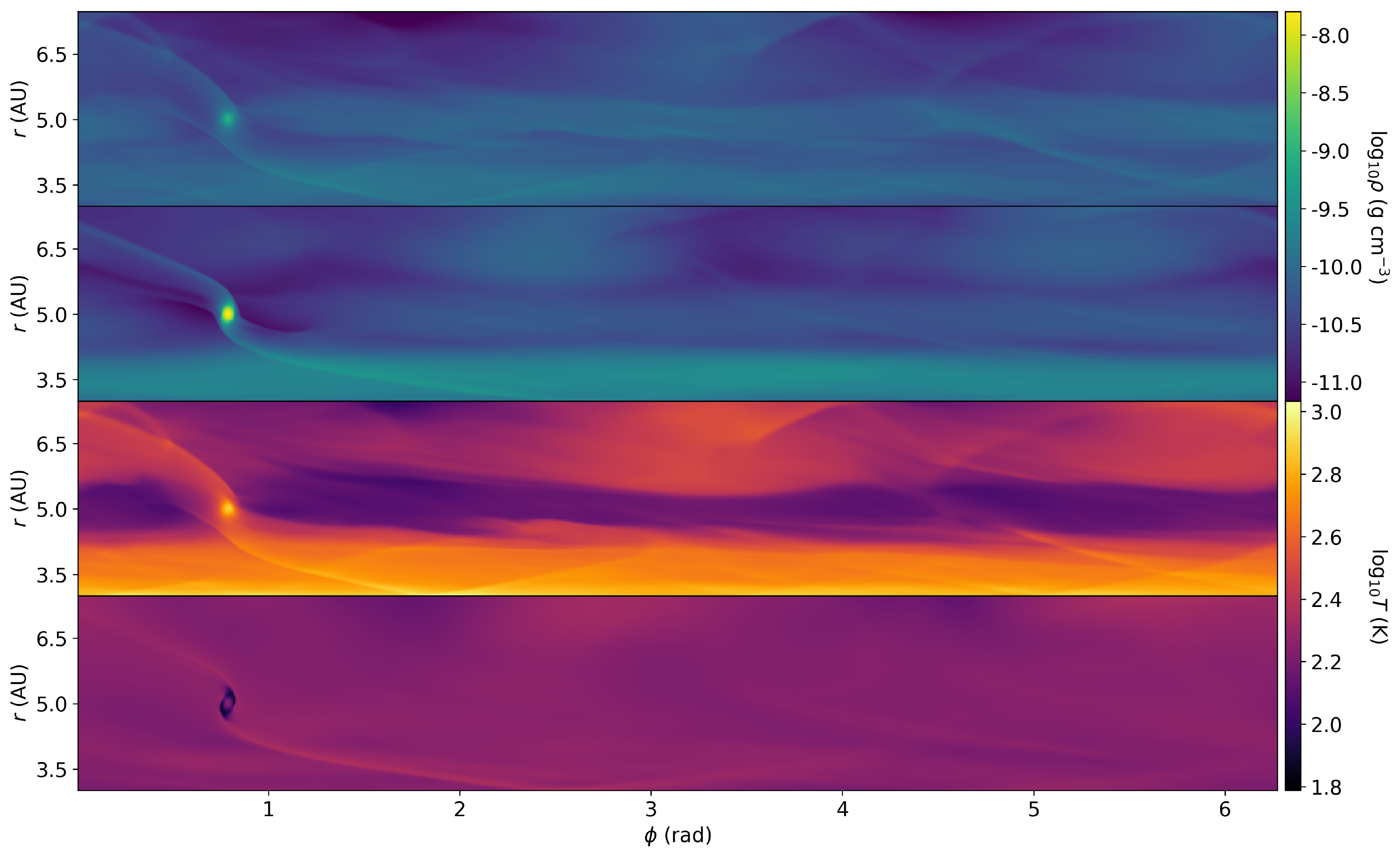}
\caption{
Horizontal $(r,\phi)$ slices in the planet-disk interaction test at 40 orbits.
From top to bottom: $\log_{10}\rho$ for runs DP\_HD
and DP\_K100, and $\log_{10} T$ for runs DP\_HD
and DP\_K100.
}
\label{fig:planetgap}
\end{figure*}

 \subsection{Planet-disk interaction}\label{SS:Planet} 

We now present an application of the code in the context
of giant planet formation. The most widely accepted
explanation for this phenomenon is the core accretion
scenario, in which giant
planets form as a consequence
of gas accretion by large ($\gtrsim$ 10 $M_\oplus$)
planetary cores in protoplanetary disks \citep{Mizuno1980,Bodenheimer1986,Pollack1996}.
The momentum exchange caused by the gravitational influence
of the protoplanet produces spiral waves in the disk,
and if the planet is sufficiently massive, i.e., if its
Hill radius exceeds the pressure scale height of the disk,
it can lead to the formation of annular gaps \citep[see, e.g.,][]{Kley2012}. These structures are affected by
the thermal structure in the disk, which consequently
affects key properties for the planet's evolution such
as its migration
and accretion rates. In
particular, the low densities produced during the formation
of gaps may produce transport of radiation from optically
thick to optically thin regions, for which the M1 closure is
particularly suited.
  
We have applied our scheme to describe the accretion process onto
a gap-opening planet embedded in a protoplanetary disk. Similar studies
have been carried out, e.g., in \cite{KlahrKley2006}, \cite{Ayliffe2012MNRAS}, and
\cite{Schulik2019}.
We consider a disk around
a solar mass star, in which a planet of mass
$M_p=M_J$ orbits at a radius $r=5$ AU.
We define this setup in a 3D domain given
in spherical coordinates as
$(r,\theta,\phi)\in[3,7.5]$ AU
$\times [\pi/2-0.12,\pi/2+0.12]\times[0,2\pi]$,
where $r=0$ corresponds to the center of mass of
the planet-star system.
The gas distribution is defined in the same way as in
Section \ref{SS:Convection}, where this time
$H_0/R_0=0.05$, $p=-3/2$, and $q=-1$. In this way,
the disk has an initially uniform $H/R$ ratio,
and the vertical domain exceeds the pressure scale
height by a factor $2.4$. In this case, no viscosity
is included. Boundary conditions
remain the same as in Section \ref{SS:Convection},
with the difference that now we apply zero-gradient
conditions for $E_r$ in the vertical boundaries
and periodic conditions to all variables in the
azimuthal direction.

The total gravitational potential is computed as a sum
of the potentials $\Phi_s$ and $\Phi_p$ due to the star
and the planet. The stellar potential is computed as
\begin{equation}
\Phi_s = 
-\frac{M_\odot G}{\vert\vert \mathbf{r}-\mathbf{r}_s \vert\vert}\,,
\end{equation}
where $\mathbf{r}_s$ is the star's position.
Following \cite{KlahrKley2006}, we compute $\Phi_p$ as
\begin{equation}
\Phi_p(\mathbf{r})=
    \begin{cases}
    -\frac{M_p G}{d_p} & \text{if } d_p \geq a_g\\
    -M_p G \left[
    \frac{d_p^3}{a_g^4}
    -2\frac{d_p^2}{a_g^3}+\frac{2}{a_g}
    \right]  & \text{if } d_p < a_g\,,
\end{cases}
\end{equation}
where $d_p=\vert\vert \mathbf{r}-\mathbf{r}_p \vert\vert$,
$\mathbf{r}_p$ is the planet's location, and
$a_g$ is a critical distance used to smooth the
potential in the vicinity of the planet.
We compute this quantity as $a_g=r_h/2$,
where $r_h$ is the planet's Hill radius, i.e.,
the approximate radius of its Roche lobe. This
quantity can be computed in terms of the reduced
mass of the system $\mu_p=\frac{M_p}{M_p+M_s}$ as
\begin{equation}
    r_h=r_p\left(\frac{\mu_p}{3}\right)^{1/3}\,,
\end{equation}
where $r_p$ is the distance between the planet and
the star. The planet's mass is smoothly incremented
during the first orbit from $0$ to $M_J$, in order
to guarantee a slow adaptation of the system and
prevent the formation of strong waves caused by
an initial nonequilibrium configuration.
In this work we do not include a local
reduction of the density per time step in the vicinity
of the planet accounting for the accretion, and focus
solely on the heating and cooling caused by radiation
transport.

We solve the Rad-HD equations in spherical coordinates
on a grid with resolution
$N_r\times N_\theta \times N_\phi = 128\times 60 \times 512$,
using $C_a=0.3$ for both radiation
and HD fields
 and $\hat{c}=c/1000$.
The grid is logarithmically spaced in the radial direction
and linearly divided in the
azimuthal direction using two
regions of different resolution,
in such a way that
the intervals $[0,\pi/2]$ and $[\pi/2,2\pi]$ have each a
resolution of $256$ zones.
We integrate these equations
in a reference frame that corotates with the planet, in
such a way that the coordinates of the latter are always
$(r_p,\theta_p,\phi_p)=(5\,\mathrm{AU},\pi/2,\pi/4)$. This
reduces the numerical diffusion around the planet, at the
cost of integrating the extra few terms that arise when
the HD equations are
transformed into this frame.
As mentioned in Section \ref{SS:RadHD},
in doing so we neglect all additional terms
arising from the transformation
of the radiation fields into
the rotating frame, which
is justified since 
$\Omega_{p}r_p/c\sim10^{-5}$,
where $\Omega_p$ is the Keplerian angular velocity of the planet.

We have run several tests with this configuration,
neglecting scattering and using
in each case $\kappa=\kappa_\mathrm{BL}$,
$\kappa_\mathrm{BL}/100$, and $\kappa_\mathrm{BL}/1000$,
where $\kappa_\mathrm{BL}$ is the Rosseland opacity
by \cite{BellLin1994} used in Section \ref{SS:Convection}.
We refer to these simulations as DP\_K1, DP\_K100, and DP\_K1000,
respectively. For comparison, we have also run a purely
hydrodynamical test with the same initial setup. We refer to this run as DP\_HD. We ran DP\_K1 and DP\_K1000 for a total
of $5.5$ orbits, where this time we define an orbit
as the Keplerian period $T_0$ at the planet's location,
while tests DP\_HD and DP\_K100 have been run for a total
of $40$ orbits.

In Figs. \ref{fig:planetedgeon} and \ref{fig:planettop}
we show, respectively, vertical and horizontal slices
showing the logarithms of $\rho$, $T$, $E_r$, and $f$
at the planet's location, taken at $t=5.5$ $T_0$.
Since in run DP\_HD we include no radiation, the $E_r$
values shown in these figures for that test correspond
to the LTE value given by $E_r=a_R T^4$. We did not
compute an $f$ value for that simulation. In each case
we overplotted the location
of the Hill sphere,
i.e., the sphere of radius $r_h$ centered on the
planet, which approximates the outer boundary of the planet's
Roche lobe.

These profiles evidence the formation two spiral arms, together with a hot gas envelope surrounding the planet that rotates in the same
direction as it. The spirals are
hotter than the surrounding material and colder than the central envelope.
Profiles
obtained in DP\_HD and DP\_K1 are almost identical, since
for high opacities the LTE limit is recovered. These
structures change and the overall temperatures decrease
for lower opacities, as the radiation begins to diffuse
away from the envelope and the spirals. Within the
Roche lobe, the maximum temperature decreases
for lower opacities from 
$543$ K in DP\_K1 ($581$ K in DP\_HD) to $280$ K in DP\_K1000.

Similar changes can be observed in the $E_r$ profiles, which
show the same structure as the temperature profiles
in DP\_HD and DP\_K1, whereas for lower opacities the energy
density begins to fill the region surrounding the planet
and the spiral arms. To see the direction of the radiative
flux, we have superimposed in these profiles
white arrows representing the value
of $\mathbf{f}=\mathbf{F}_r/E_r$, using the same scale
for every run. Together with the $f$ plots,
these profiles evidence the different regimes of
radiation transport in the different runs. In DP\_K1
the value of $f$ remains below $0.07$, and radiation
is entirely in the diffusion regime. In DP\_K100
we begin to see radiation transported away from the spiral
arms with a maximum $f$ of $0.25$. On the other hand, the
vertical slices show vertical transport of radiation
at a maximum $f$ of $0.14$ through the low-density regions
above and below the planet, which were caused by the
planet's gravitational attraction. Run DP\_K1000, on the
other hand, shows a transition between the diffusion regime,
observed within the envelope and the spirals, and the almost freely transport streaming away from the spirals and in the
vertical direction, with maximum $f=0.94$. 
At this time, the radial optical depth across the Hill sphere is approximately
$28400$, $690$, and
$150$ in DP\_K1,
DP\_K100, and DP\_K1000
respectively,
whereas the vertical
optical depth
across the Hill sphere
in each of these cases
is of $19000$, $420$, and $50$. In DP\_K100 and DP\_K1000, the observed
radiative losses occur
despite these high values
since most of this optical
depth is caused by the
large accumulation of 
mass close to the planet's location,
whereas diffusion
is still possible
around this region.
At $t=0$, the radial
optical depths
across the same region
are $3040$, $30$, and
$3$ in DP\_K1,
DP\_K100, and DP\_K1000
respectively,
while the vertical
ones are $2100$, $21$, and $2$.

For decreasing opacities, the lower pressure support
caused by radiation diffusion allows for a larger
infall of matter onto the planet.
This produces larger maximum densities
in the envelope and also lower densities
above and below the planet,
as shown in the top rows of Figs. \ref{fig:planetedgeon} and \ref{fig:planettop}.
At that time, maximum densities range from
$3.8\times 10^{-9}$ g cm$^{-3}$ in DP\_K1
($3.5\times 10^{-9}$ g cm$^{-3}$ in DP\_HD)
to $2.1\times 10^{-9}$ g cm$^{-3}$ in DP\_K1000.
In the same plots, we have overplotted with
white arrows the gas velocity in the planet's
corotating frame, using the same scale for
every run. In the vertical profiles, it can
be seen that matter is transported into the
envelope predominantly from the poles,
with maximum vertical mass fluxes ranging from
$1.58\times 10^{-5}$ g cm$^{-2}$ s$^{-1}$ in DP\_K1
($1.46\times 10^{-5}$ g cm$^{-2}$ s$^{-1}$ in DP\_HD)
to $1.24\times 10^{-4}$ g cm$^{-2}$ s$^{-1}$ in DP\_K1000. In the horizontal profiles, we notice
that conservation of angular momentum
in the envelope causes the latter to rotate faster for decreasing opacities, with maximum
angular velocities corresponding to rotational
periods of $515$ days in DP\_K1 ($655$ days in
DP\_HD) and $186$ days in DP\_K1000.

\begin{figure}[t]
\centering
\includegraphics[width=0.47\textwidth]
{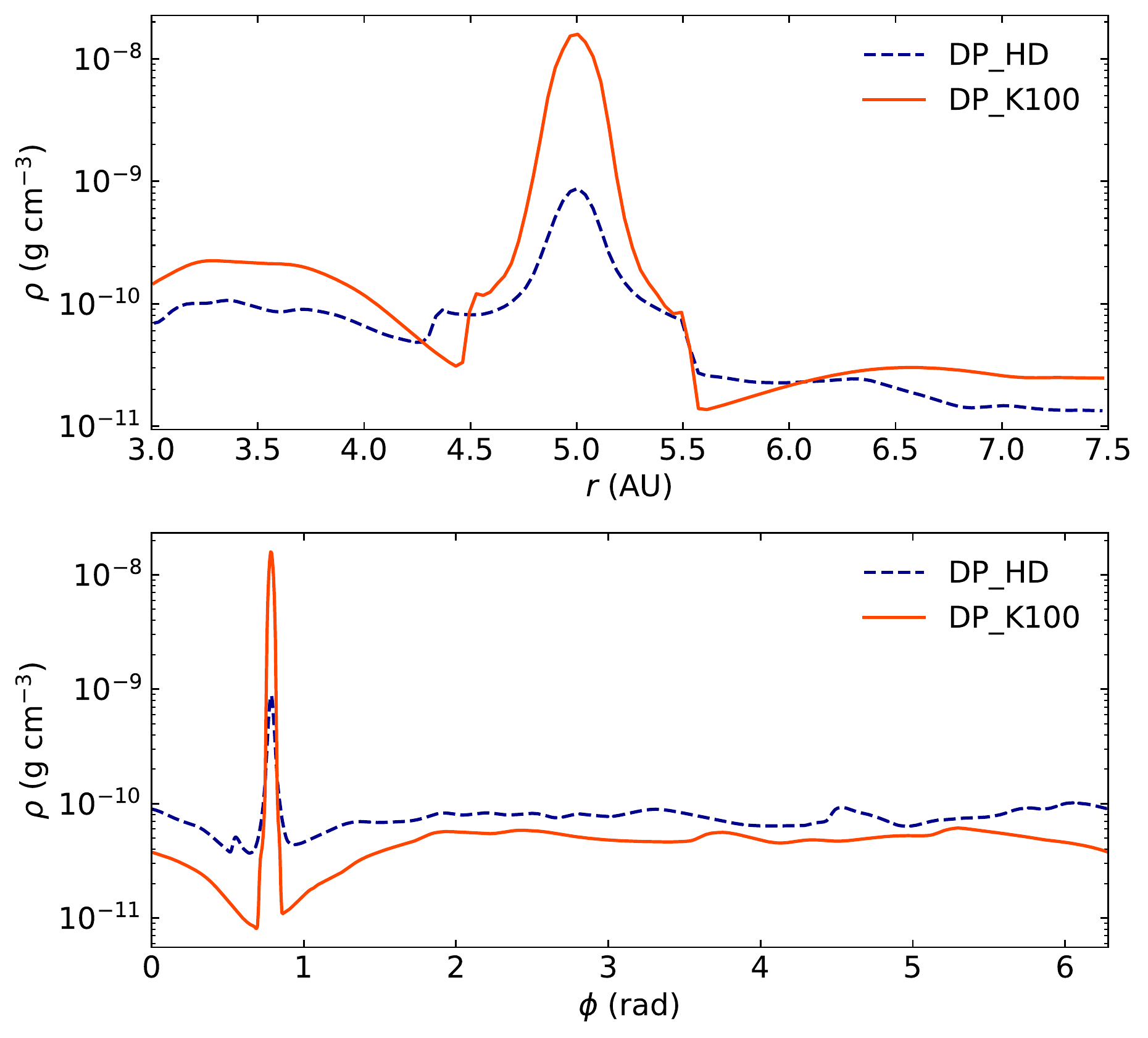}
\caption{1D mass density profiles at 40
orbits in runs DP\_HD and DP\_K100,
computed as a function of $r$ for
fixed $\phi=\phi_p$ and
$\theta=\theta_p$ (top), and as a function
of $\phi$ for fixed
$\theta=\theta_p$ and $r=r_p$ (bottom).
}
\label{fig:planet1dprofiles}
\end{figure}

\begin{figure}[t]
\centering
\includegraphics[width=0.47\textwidth]
{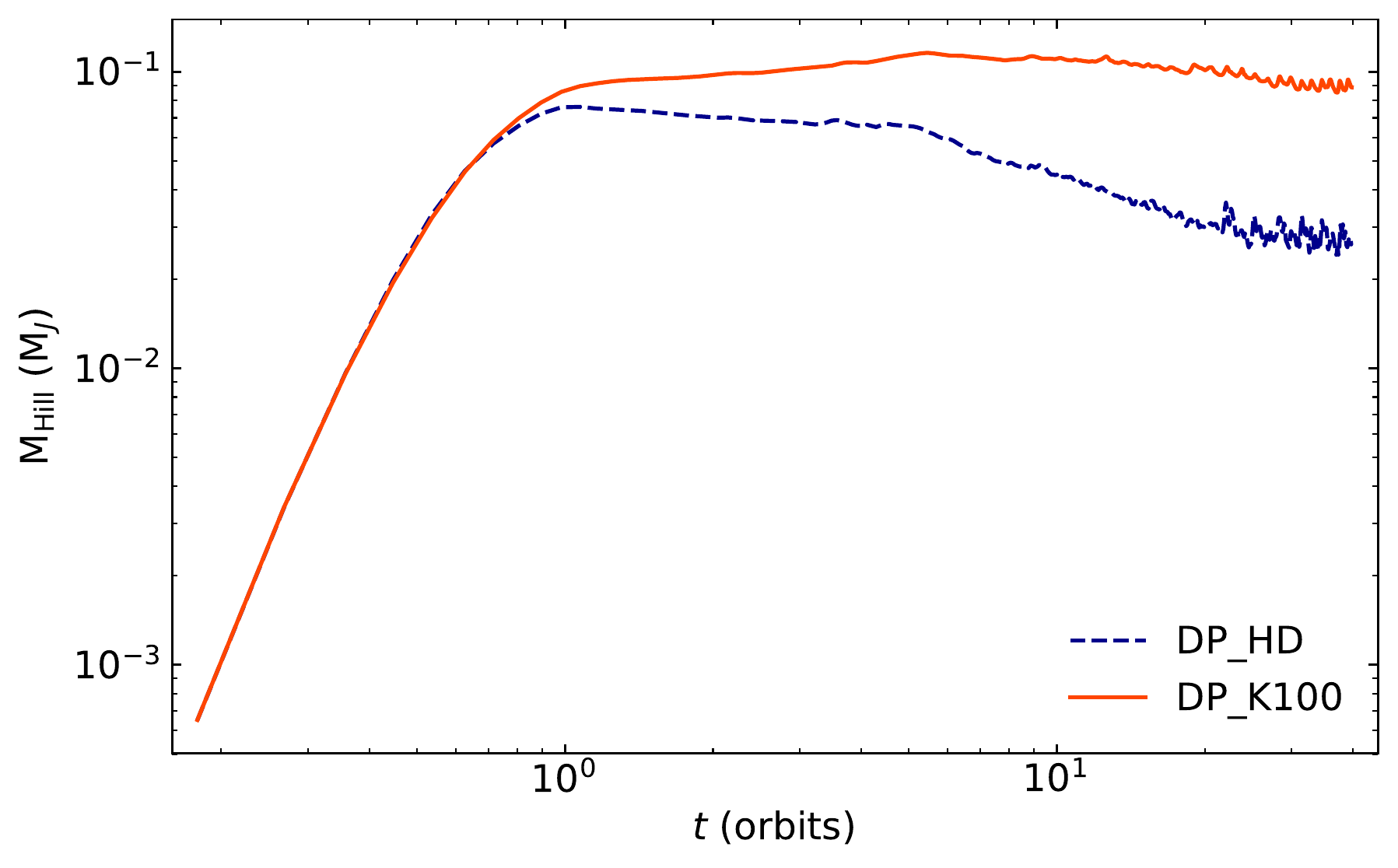}
\caption{Mass enclosed in the Hill sphere
in runs DP\_HD and DP\_K100 as a function
of time.}
\label{fig:Mhill}
\end{figure}

In Fig. \ref{fig:planetgap} we show $(r,\phi)$
profiles at $z=0$ for runs DP\_HD and DP\_K100
after $40$
orbits. Again, we observe lower temperatures and
larger maximum densities close to the planet
in DP\_K100. We can see that the temperature 
distribution is much more uniform in DP\_K100
than in DP\_HD, where the temperature decreases
in a neighborhood of the planet radius in the
entire domain. We also notice structural
differences in the gas density distribution,
where matter within the planet's horseshoe orbit
has a lower density in DP\_K100 than in DP\_HD.
This can be clearly seen in Fig.
\ref{fig:planet1dprofiles}, in which we
show the gas density along the radial and azimuthal
directions at the planet's location. The first of
these plots shows that the density in DP\_K100
is larger than in DP\_HD away from the planet
except at a distance of $\sim 2 r_h$ from the
planet's location, where the density in DP\_K100
presents a sharp decrease unobserved in DP\_HD.
It is likely in this case that the vertical
shrinking of the disk caused by
radiative diffusion favors a faster formation of a gap at $r\sim5$ AU
when compared to DP\_HD.

We computed as
a function of time the total mass $M_\mathrm{Hill}$
within the Hill sphere in both simulations, shown
in Fig  \ref{fig:Mhill}. In run DP\_K100, $M_\mathrm{Hill}$ exceeds its value in DP\_HD
from the first orbit, reaching after 40 orbits
$0.088$ $M_J$ in DP\_K100 and $0.026$ $M_J$ in DP\_HD.
This shows that reducing the opacity would lead in
this case to a faster growth of the planet. Similar conclusions are reached, e.g., in \cite{Movshovitz2010} and \cite{Schulik2020}.
We intend to carry high-resolution studies of this
problem in the near future, using better estimates for the
Rosseland and Planck opacities and including the
mass decrease caused by accretion onto the planet.

\subsection{Stellar irradiation}\label{SS:Pascucci} 

We  tested the implementation of the irradiation terms
by reproducing the benchmark by \cite{Pascucci2004},
which consists in computing the equilibrium temperature
of a static disk irradiated by a central star. We 
compared temperature
distributions obtained
with both the presented module and the Monte Carlo radiative transfer code \sftw{RADMC-3D} \citep{Dullemond2012}.
In both cases, the gas density is
defined in spherical coordinates
$(r,\theta)\in [1,1000]\,\mathrm{AU}\times [0,\pi]$
as
\begin{equation}
    \rho(R,z) = \rho_0  \left( \frac{500\, \mathrm{AU}}{R} \right) \exp \left(
    -\frac{\pi}{4}\left(
    \frac{z}{h(r)}
    \right)^2
    \right)\,,
\end{equation}
where $(R,z)=r\,(\cos\theta,\sin\theta)$ and
$h(R)=125\,\mathrm{AU}\times(R/500\,\mathrm{AU})^{1.125}$. 
To compute the opacities for both irradiation and
radiation-matter interaction terms, we use the frequency-dependent absorption
cross sections by \cite{DraineLee1984},  derived for silicate
dust particles with sizes between $0.003$ and $1$ $\mu$m.
To convert the tabulated cross sections into opacity 
coefficients, we assume the dust grains to have
a radius of $0.12$ $\mu$m and a density of $3.6$ g cm$^{-3}$.
We set $\rho_0=6.66\times10^{-17}$ g cm$^{-3}$ and a uniform dust-to-gas mass ratio of $0.01$, in such a way that the absorption optical depth at $550$ nm for a radial path that crosses the domain along the midplane equals $\tau=100$.

In the Rad-HD simulation, the irradiation flux
is computed as
\begin{equation}\label{Eq:Firrad}
    \mathbf{F}_\mathrm{Irr}(r,\theta) =
    \pi\left(\frac{R_s}{r}\right)^2
    \int_{\nu_\mathrm{min}}^{\nu_\mathrm{max}} \mathrm{d}\nu\,
    B_\nu(T_s)\,
    e^{-\tau(r,\theta,\nu)}\, \hat{\mathbf{r}}\,,
\end{equation}
where $T_s= 5800$ K is the star temperature,
$R_s=R_\odot$ is the star radius,
$B_\nu(T_s)$ is the Planck radiative intensity, and
$[\nu_\mathrm{min},\nu_\mathrm{max}]=
[1.5\times 10^{11},1.5\times 10^{15}]$ Hz
is the considered frequency range.
% The factor $\pi$ on the right-hand side of this equation comes from integrating the projected radiative intensity onto the direction of propagation, in such a way that, in absence of opacity, $\mathbf{F}_\mathrm{Irr}(r) = (R_s / r)^2 \sigma_\mathrm{SB} T^4 \hat{\mathbf{r}}$.
 The optical depth is computed along radial trajectories as
\begin{equation}
    \tau(r,\theta,\nu) = 
    \int^r_{1\,\mathrm{AU}}\mathrm{d}r'\,
    \kappa(\nu)\,\rho(r',\theta)\,,
\end{equation}
where $\kappa(\nu)$ is the tabulated frequency-dependent
absorption opacity, while scattering is neglected.
In the radiation-matter interaction
terms (Eq. \eqref{Eq:Gcomov}), we compute $\kappa$ and
$\chi$ respectively as their Planck and Rosseland means
evaluated at the local gas temperature.

We integrate the evolution equations of radiation
fields and gas energy neglecting the advection
terms of the latter, namely,
\begin{equation}
    \frac{\partial E}{\partial t}
    = c G^0 - \nabla\cdot \mathbf{F}_\mathrm{Irr}\,.
\end{equation}
The gas and radiation energy densities are initially set
at LTE at a temperature of $10$ K in the entire domain.
We solve the resulting system of equations on a 2D
spherical grid of resolution $N_r\times N_\theta =240\times100$
increasing logarithmically in the radial direction,
using the same boundary conditions for the radiation fields as
in Section \ref{SS:Convection} and $\hat{c}=c/100$. 
The same grid is used in the \sftw{RADMC-3D} Monte Carlo computation. In that case, the trajectories of $10^{10}$
photon packages are tracked and used to compute the disk temperature taking into account the full frequency
dependency of the dust opacity. The photons are injected
at $r=0$ with an energy distribution proportional
to $B_\nu(T_s)$, and normalized in such a way that the
total luminosity equals that of an emitting spherical blackbody
with radius $R_s$ and
temperature $T_s$.

\begin{figure}[t]
\centering
\includegraphics[width=0.47\textwidth]
{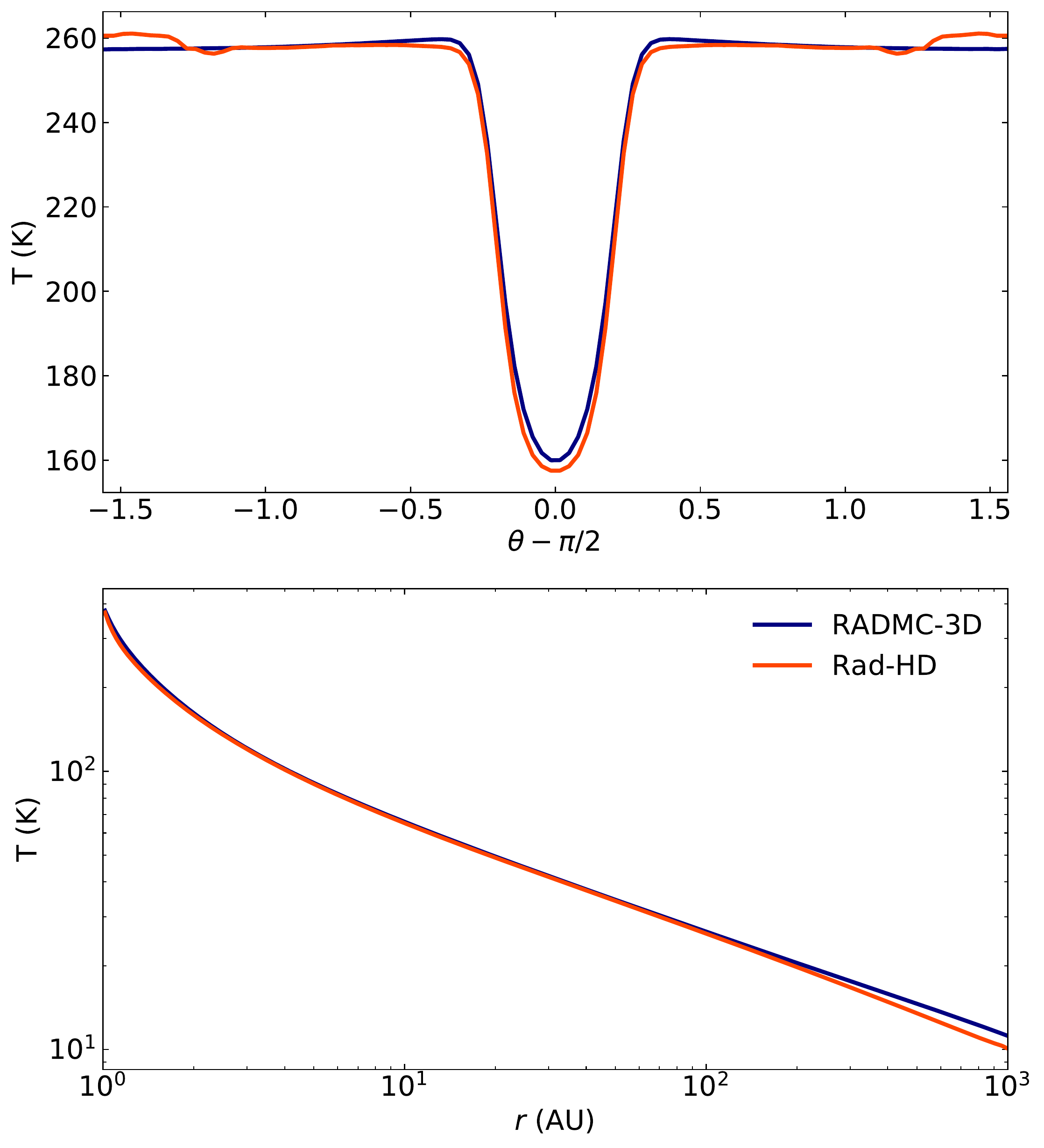}
\caption{
Temperature distributions in the stellar
irradiation test obtained with the
presented Rad-HD module (orange) and the Monte Carlo
radiative transfer code \sftw{RADMC-3D} (blue),
shown at $r=2$ AU (top) and at the disk midplane (bottom).
}
\label{fig:IrradPascucci}
\end{figure}

In Fig. \ref{fig:IrradPascucci}, we show 1D slices of the
resulting temperature profiles for both simulations, shown
as a function of $\theta-\pi/2$ at $r=2$ AU and as a function of
$r$ at the disk midplane. In the first case, both temperature
distributions show a good agreement, with relative differences
of under $3\%$ of their values. We note that the temperature obtained with Rad-HD exceeds that computed with \sftw{RADMC-3D} close to the azimuthal boundary. This feature is caused by an energy accumulation
originated by converging fluxes onto
the vertical axis, and disappears if a smaller polar extent is chosen. We obtain in both simulations that the midplane temperature decreases  approximately as $r^{-0.4}$ for $r>10$ AU.
The difference between the radial temperature profiles stays below $5\%$ between $1$ and $2$ AU, remains under $1\%$ between $2$ and $70$ AU, and steadily grows up to its maximum value of $10\%$ at
$1000$ AU. At that radius, this percentage represents an absolute
difference of $1.1$ K, and in fact we have verified that this difference stays below $1.2$ K for $r>2.5$ AU. Overall, we observe
a good agreement between both solutions, comparable for
instance with that obtained in \cite{Flock2013} and \cite{MignonRisse2020}. 

%% file: summary.tex
\newpage
 \section{Conclusions}\label{S:Conclusions}
 
  %Future: planet-disk, AMR, open source
 % M1. Beaming through gap
 
% This work: same values of Rosseland and Planck (Malygin 2014). Future studies need to include accurate... 
The goal of this paper was to develop a Rad-HD scheme of general
application
that is optimized for studies of accreting planets in circumstellar disks. We chose the M1 scheme for this approach as it can handle the anisotropy of the radiation field around an accreting planet and specifically the expected accretion shock.

We have presented a radiative transfer
module integrated within the
HD module of the \sftw{PLUTO} code.
The code solves the evolution equations of HD and radiative fields
separately through operator
splitting, applying substepping for
the evolution of radiation fields
in order to reduce the overall computational cost. The number of
radiation substeps is reduced by
applying the RSLA, and two
different IMEX-Runge Kutta schemes can be applied within
each substep to integrate the radiation
advection and interaction terms.
Among other solvers, we have
implemented the HLLC Riemann solver
for radiation transport introduced
in \cite{MelonFuksman2019} in the
context of Rad-RMHD. The code has
been adapted to all available geometries included in \sftw{PLUTO},
is fully parallel, and can be 
implemented in rotating frames
provided that the
relativistic corrections
to the radiation fields
when transformed into such frame
are negligible,
which is particularly useful in
global simulations of 
circumstellar disks and planetary
accretion. 

We have tested the code in different scenarios relevant to the physics of
protoplanetary disks,
paying particular attention to the 
behavior of the solutions when
different values of the speed
of light are chosen. In the considered radiative shocks benchmarks,
we observe that subcritical
shock solutions are accurate
in a broader $\hat{c}$ range
than supercritical shocks.
The obtained solutions
with $\hat{c}=c$ are in agreement
with those reported in other works.
We have estimated the energy loss caused by the RSLA when energy
is introduced into the system
from the domain boundaries,
obtaining approximate lower bounds
to the value of $\hat{c}$.
On the other hand, all runs of the 1D vertical diffusion test in a static
disk yield energy and flux distributions that converge
to the exact stationary
solution in different
timescales. We observe slight
deviations with respect to the
exact solution caused by
operator splitting error, that
get reduced for decreasing $\hat{c}$.

We have applied the code in 2D
simulations of viscously heated protoplanetary disks. 
The obtained solutions are almost
indistinguishable for $\hat{c}$
values larger than the theoretical
limit obtained by applying the
validity conditions for the RSLA
given in \cite{Skinner2013}, and
are clearly different for lower
values.
We obtain that the mean convective
and radiative heat fluxes in the
vertical direction are reduced 
for decreasing $\hat{c}$. We also
compare these effects in terms of the time-averaged Nusselt number, whose maximum value decreases when $\hat{c}$ is reduced.

We ran 3D HD and Rad-HD
simulations of the gas accretion
by a giant Jupiter mass core
embedded in a protoplanetary
disk. We computed the
joint evolution of gas and
radiation for three different
opacity regimes, observing
in every case
the formation of spiral arms
and a hot rotating gas envelope surrounding the planetary core.
For the
highest employed opacity, the LTE
limit is recovered and the solutions
are almost identical to those obtained with HD.
For lower opacities,
the produced envelope becomes more compact due to the
lower pressure support caused by
radiative losses and rotates
faster due to
conservation of angular momentum. 
In such cases, a transition between the diffusion and almost free-streaming
regimes is observed as radiation
is transported away from the envelope
and the spirals.
%For the intermediate
%opacity value, radiation
%begins to be diffused horizontally %from the spiral arms and vertically
%through the low-density regions above and below the planet formed by its gravitational attraction.
%For the lowest opacity, a transition
%between the diffusion and almost
%free streaming regimes can be seen
%as radiation is transported away 
%from the spirals and in the vertical
%direction.
After $40$ orbits, the simulation
with the intermediate opacity value
shows a sharper gap at the planet location and overall lower temperatures than in the HD
adiabatic case. We have computed
the total mass inside the planet's
Roche lobe as a function of time,
showing higher values in the Rad-HD
case, which could indicate a faster
planet growth for decreasing opacity.

We have further studied the
performance of our scheme in
standard tests for comparison
with other methods. We have
verified the accuracy of the
IMEX-SSP2(2,2,2) method, which shows
a convergence order closer to 2
than the operator-split scheme by \citet{Skinner2013}.
We have studied the parallel performance of the code in
2D and 3D setups
using up to $1280$ processors, 
in which case we obtain
efficiencies of 93\%
in 2D and 85\% in 3D.
Future developments of this module
will include the implementation of
the adaptive mesh refinement routines
already present in \sftw{PLUTO}.
The module presented in this work will be included in forthcoming releases of
\sftw{PLUTO}, which can be downloaded
from
\url{http://plutocode.ph.unito.it/}.

Future studies of our M1 Rad-HD scheme will
expand on the modeling of gas accretion
onto planetary cores, the use of
realistic Rosseland and Planck opacities, and higher resolutions
achieved through adaptive mesh
refinement. 
Currently we are comparing our results on the temperature structure around the planet and the intensity of radiation with detailed  Monte Carlo continuum radiative transfer simulations \citep{Krieger2020}, in a collaboration on deriving the characteristics of exoplanets from observations of for various current and future instruments including ALMA \citep{2002Msngr.107....7K}, PIONIER \citep{2011A&A...535A..67L}, and MATISSE \citep{2014Msngr.157....5L}.

%% file: performance.tex
\section{Performance tests}\label{S:Performance}
\subsection{Damped linear waves}\label{SS:LinearWaves} 
 
We tested the convergence rate of the implemented IMEX schemes
by investigating the evolution of damped linear radiation waves in
a static absorbing medium. We have reproduced the setup by \cite{Skinner2013}, in which the material's emission
is neglected. This leads to
the following evolution equations for the radiation quantities:
\begin{equation}
\begin{split}
    \frac{1}{\hat{c}}\frac{\partial E_r}{\partial t}
    +\nabla\cdot \mathbf{F}_r &= - \kappa \rho E_r \\
    \frac{1}{\hat{c}}\frac{\partial \mathbf{F}_r}{\partial t}+\nabla\cdot \mathbb{P}_r &= - \kappa \rho F^x_r\,.
\end{split}
\end{equation}
We define the initial condition as
\begin{equation}
    \mathcal{U}_r(\mathbf{r},0) =
    \mathcal{U}_0(\mathbf{r}\cdot\mathbf{n})
    =
    \left[
    E_0+\varepsilon\sin \left(
    \frac{2 \pi}{\lambda}\, \mathbf{r} \cdot \mathbf{n}
    \right)
    \right]
    \left(
    \begin{array}{c}
    1 \\
    \mathbf{n}
    \end{array}
    \right)\,,
\end{equation}
where $\mathbf{n}$ is a unit vector indicating the direction of the radiative flux and the $E_0$, $\epsilon$, and $\lambda$ parameters correspond, respectively, to the mean value, the amplitude, and the wavelength of the initial state. This initial condition satisfies
$\vert\vert\mathbf{F}_r\vert\vert=E_r$, and therefore the pressure tensor is proportional to $E_r$ as $\mathbb{P}_r=E_r\,\mathbf{n}\,\mathbf{n}$ (see Section \ref{S:Equations}). The exact solution of this initial value
problem is a damped wave of the form
\begin{equation}
    \overline{\mathcal{U}}_r(\mathbf{r},t) = 
    \mathcal{U}_0(\mathbf{r}\cdot\mathbf{n}-\hat{c}\,t)
    \,e^{-\rho\kappa \hat{c} t}\,,
\end{equation}
which consistently maintains the free streaming condition $\vert\vert\mathbf{F}_r\vert\vert=E_r$ throughout its entire
evolution. We parameterize the direction of propagation as
$\mathbf{n}=\left(
\cos\alpha \cos\beta , \cos\alpha \sin\beta, \sin\alpha
\right)$\,,
with $\alpha\in[0,\pi]$ and $\beta\in[0,2\pi)$.

We have computed the evolution
of $\mathcal{U}_r$ using
the IMEX1 and IMEX-SSP2(2,2,2) methods (see
Section \ref{SS:RadStep}).
Simulations were run in in 1D, 2D, and 3D in each case,
using the HLL and HLLC Riemann solvers. We chose the parameters $E_0=1$, $\varepsilon=10^{-6}$, $\lambda=1$, and $\hat{c}=c=1$. 
 We conducted in each case a resolution study using uniform Cartesian grids with periodic boundary conditions in every direction. The employed resolution is parameterized with an integer $N$ in the range $[2^4,2^8]$. We use the domains $[0,1]$, $[0,\sqrt{5}]\times[0,\sqrt{5}/2]$, and $[0,3]\times[0,3/2]\times[0,3/2]$ and the angles 
 $(\alpha,\beta)=(0,0)$, $(0,\tan^{-1}(2))$, and
 $(\tan^{-1}(2/\sqrt{5}),\tan^{-1}(2))$ in 1D, 2D, and 3D
 respectively. In this way, the domain length in each direction corresponds to one wave period. The time step is set as in Eq. \eqref{Eq:CourantRad}, with $C_a=0.3$.

\begin{figure*}[t!]
  \centering
  \includegraphics[width=0.5\linewidth]{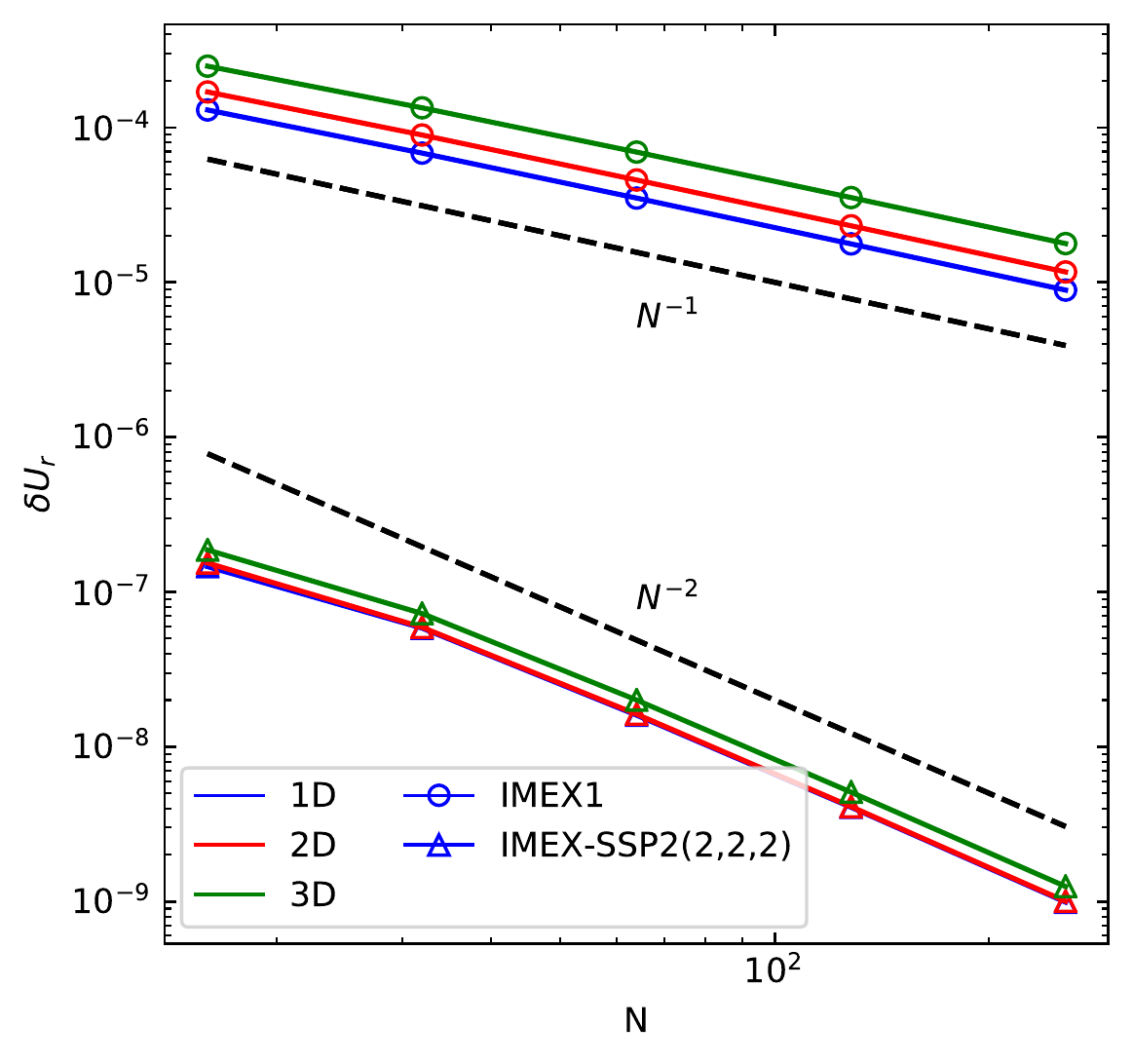}
\caption{2-norm of $\delta \cU_r$
for the 1D, 2D, and 3D damped wave test (blue, red, and green solid lines respectively) as a function of the resolution parameter $N$.
The circle and triangle symbols correspond to computations carried
out with the IMEX1 and IMEX-SSP2(2,2,2) methods, respectively. Dashed black lines show the ideal convergence slope for each method.}
\label{fig:linwaveres}
\end{figure*}

For each integration
method and resolution, we compute the
$L_1$-difference
$\delta \cU_r$ between the obtained
$\cU_r$ and the exact solution at $t=1$, i.e, after one period,
defined as
\begin{equation}
    \delta \cU_r = \frac{1}{2 N^d}
    \sum_{i,j,k}
    \vert
    \cU_{r,i,j,k} - \overline{\cU}_{r,i,j,k}
    \vert_1\,,
\end{equation}
where $\vert\cdot\vert_1$ denotes the $L_1$ norm, $d$ is the problem's dimension,
and the indices $(i,j,k)$ run over
all grid cells.
The obtained values of $\delta \cU_r$
are shown in Figure \ref{fig:linwaveres}
as a function of $N$.
In each case, the numerical solutions converge to the exact ones at the expected rate, i.e., $1$ for IMEX1 and $2$ for IMEX-SSP2(2,2,2).
The errors computed with the latter are comparable to those reported
by \cite{Skinner2013}, while the
IMEX-SSP2(2,2,2) method is closer
to order 2 accuracy.
It is remarkable that the errors computed with IMEX-SSP2(2,2,2)
are around 3 orders of magnitude smaller than with IMEX1, since
the former method computes the mean value of the wave much
more accurately than the latter.
Unlike in the Riemann shock tests in \cite{MelonFuksman2019}, in this case we observe no difference between the accuracy of the solutions computed with the HLL and HLLC solvers, since no contact waves are created when radiation transport occurs in only one direction.

\subsection{Marshak wave}\label{SS:Marshak} 

The Marshak wave test, named after the work by \cite{Marshak1958}, is a
radiative transfer
problem generally used as a
standard benchmark for Rad-HD codes
that studies the propagation of a planar radiation front
into a purely absorbing, cold, homogeneous medium.
In this setup, radiation is injected
from the left boundary of a 1D domain
defined as $x\geq 0$.
A semi-analytic solution of this problem is given
in \cite{SuOlson1996} under the diffusion and Eddington
approximations, i.e., assuming the validity of Eq.
\eqref{Eq:Diff}, and assuming constant opacity. Additionally,
as proposed by \cite{Pomraning1979}, it
is assumed as a simplification
that the constant-volume heat capacity $c_v$ of the material
is proportional to $T^3$, where
$c_v=\partial (\rho \epsilon)/\partial T$
(see Eq. \eqref{Eq:GasEnergyDensity}). Taking $\mathbf{v}=\mathbf{0}$, this is
equivalent to redefining the gas temperature in such
a way that $E \propto T^4$.

We have approached this problem by
solving the Rad-HD equations with constant $\rho=1$ and
null velocity, taking $c=\hat{c}=a_R=\kappa=\rho=1$
and $E=T^4$. In the notation used by \cite{SuOlson1996},
the latter choice corresponds to setting $\epsilon=1$. Unlike
in that work, we do not use
the diffusion and Eddington approximations, and instead compute the radiation flux by means of the last of Eqs.
\eqref{Eq:RadHD}.

\begin{figure*}[t!]
  \centering
  \includegraphics[width=\linewidth]{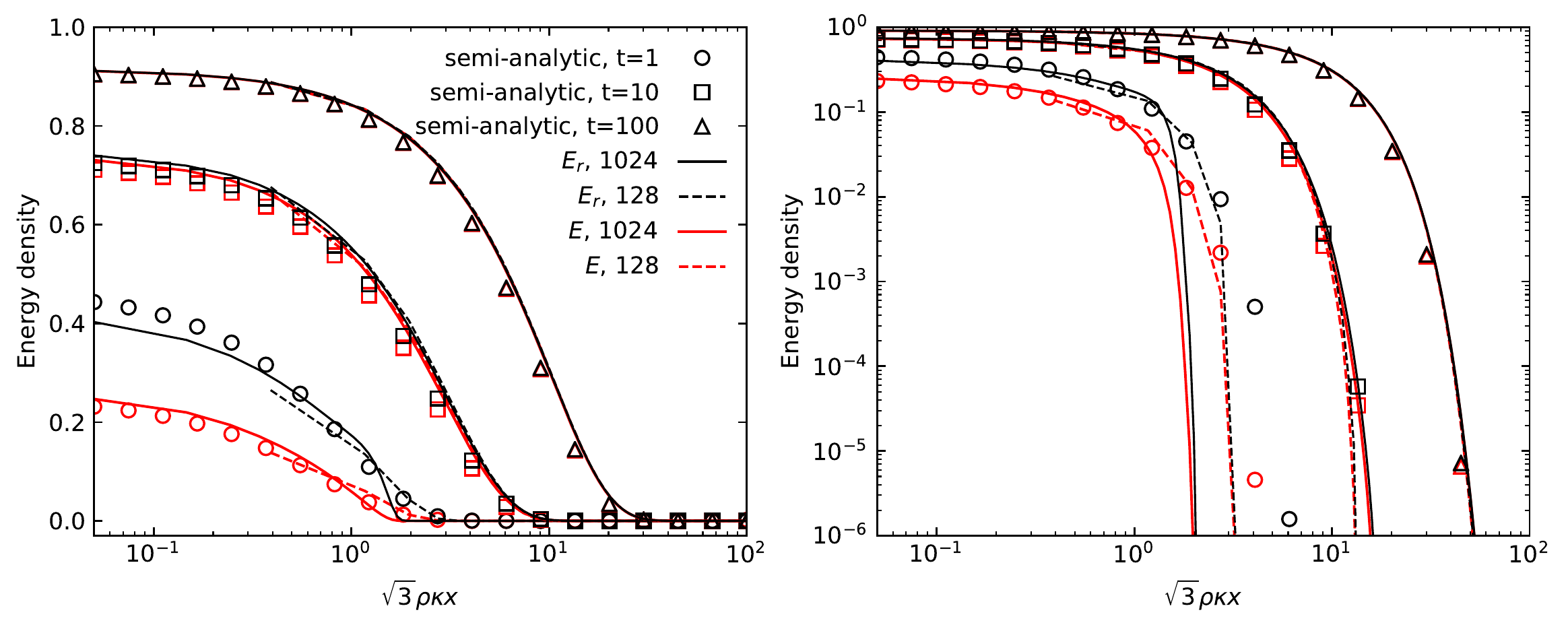}
\caption{Gas (red) and radiation (black) energy densities
obtained in the Marshak wave test
at $t=1$, $10$, and $100$,
represented on semi-log (left)
and log-log (right) scales
as a function of
$\sqrt{3}\rho\kappa x$.
Solid and dashed lines correspond
to numerical values obtained at
resolutions of $1024$ and $128$
zones, respectively.
The semi-analytical solutions
by \cite{SuOlson1996} are
shown with circle, square,
and triangle symbols at $t=1$, $10$, and $100$ respectively.
}
\label{fig:marshak}
\end{figure*}

For a better comparison with other works, we define the computational domain as $x\in [0,100/\sqrt{3}]$.
In this way, the total optical depth
of the domain is $100/\sqrt{3}\approx 57.7$.
We initially set uniform gas and
radiation energy densities as
$E_r=E=10^{-8}$, while $F^x_r=0$.
These same relations are also
imposed for $t>0$ at the right boundary, while on the left one we use the Marshak boundary condition given by
\begin{equation}\label{Eq:Marshak}
    E_r + 2F^x_r = 4 F_\mathrm{inc}\,,
\end{equation}
where $F_\mathrm{inc}=1/4$ is the
flux incident on the $x=0$ surface.
This condition is imposed by computing $E_r$
at $x=0$ using the semi-analytical
solution by \cite{SuOlson1996},
and subsequently using 
Eq. \eqref{Eq:Marshak}
to compute $F^x_r$.
We employ the IMEX1 method with
the HLLC solver and the second-order linear TVD
Van Leer reconstruction
scheme, with $C_a=0.4$.

The obtained values for $E$ and $E_r$
are shown in Fig. \ref{fig:marshak}
at $t=1$, $10$ and $100$
at the resolutions of 128 and 1024
zones, together with their semi-analytical values.
In each case, the left boundary 
condition creates a freely streaming
radiation front that propagates into
the domain while transitioning into
the diffusion regime as it interacts
with increasingly large amounts of matter.
At $t=1$, the reduced radiative
flux reaches $f=1$ at the wave front,
while at $t=100$ this value is reduced
to $0.25$,
which corresponds to
$\xi\approx 0.36$ (see Eqs. \eqref{Eq:M11}--\eqref{Eq:M13}). In the same way, the
radiation and gas energy densities 
are largely different at $t=1$ and
almost identical at $t=100$,
since they are both
equal to $T^4$ in LTE.

As expected, the numerical
solutions approach the 
semi-analytical ones as the diffusion
regime is reached. 
The agreement between both solutions
is comparable to that obtained in 
\cite{Gonzalez2007} and
\cite{Skinner2013}.
The obtained solutions are similar to those shown in \cite{Skinner2013} with the same chosen parameters and
at the same resolutions. However,
as in \cite{Gonzalez2007}, we still
observe at later times a difference
between the semi-analytic and
numerical solutions that is not
apparent in \cite{Skinner2013},
possibly due to the different
operator splitting scheme used in that work.
Such a difference is however 
expectable, since the wave
front is outside the diffusion
regime through almost its
entire evolution.

 \subsection{Parallel performance}\label{SS:Parallel} 
 
 \begin{figure}[t]
\centering
\includegraphics[width=0.47\textwidth]
{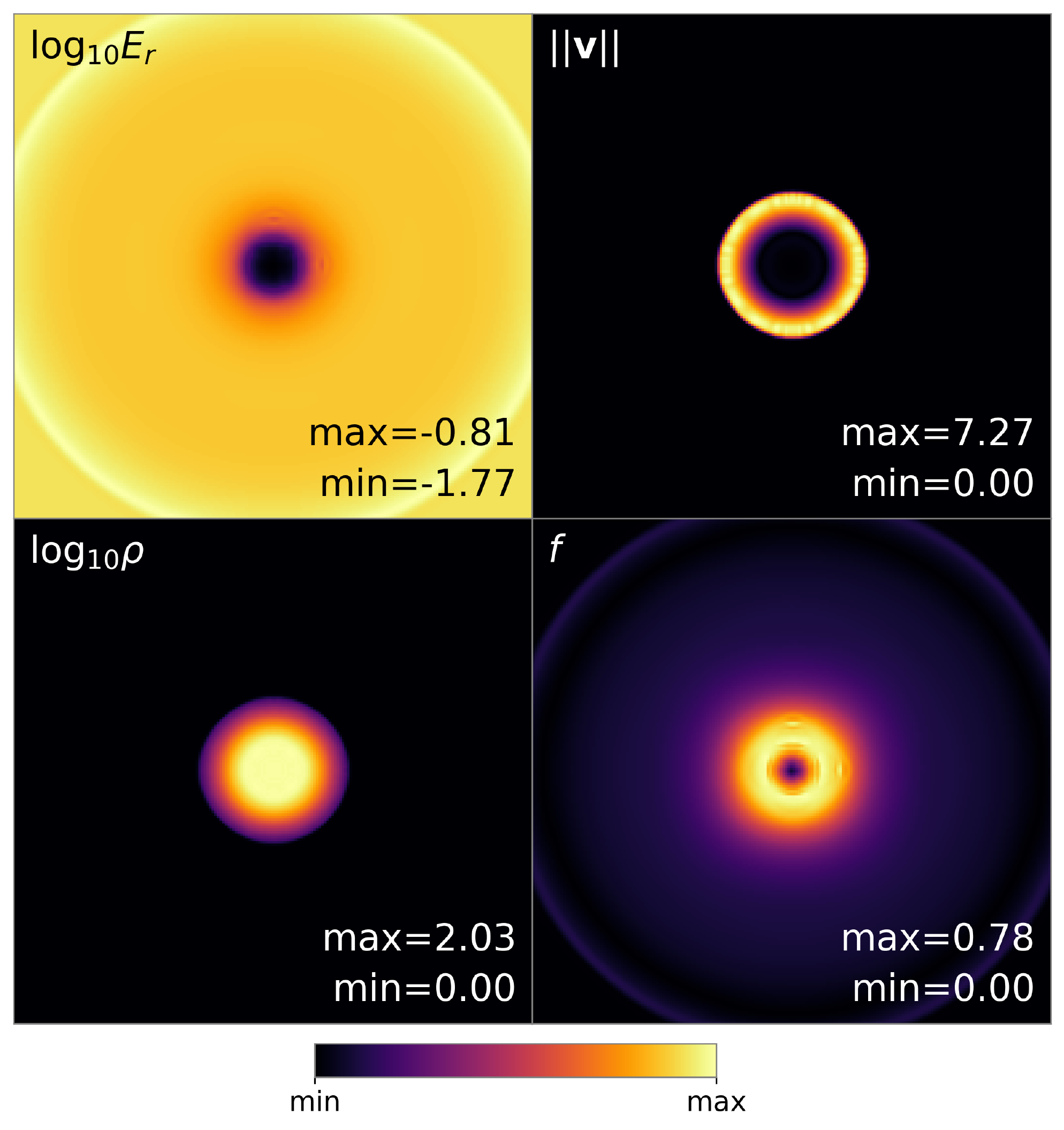}
\caption{2D slices at $z=0$ of the
3D blast wave test used for the parallel scaling analysis,
showing $\mathrm{log}_{10}E_r$, $\mathrm{log}_{10}\rho$, $\vert\vert\mathbf{v}\vert\vert$,
and $f$ at $t=0.007$.}
\label{fig:blast}
\end{figure}
 
 We tested the parallel 
 scalability of the presented
 code in strong scaling through
 2D and 3D computations.
 With this purpose, we 
 set up a configuration in which
 a blast wave is created from an
 overpressurized region of radius
 $R_0=0.1$ in the center of a cubic
 domain of side length $L=1$.
 All fields are
 initially uniform both outside
 and inside of this region, with
 $\rho=p_g=100$ inside and $\rho=p_g=1$ outside.
 Both $\rho$ and $p_g$ decrease
 linearly from their maximum to their minimum values between $r=0.08$
 and $0.1$, where
 $r=\sqrt{x^2+y^2+z^2}$
 ($r=\sqrt{x^2+y^2}$) in 3D
 (2D).
 Initial LTE is imposed
 in the entire domain, with
 $a_R=\mu m_p/k_B=1$ and $\Gamma=1.4$. 
 We set as well
 $\kappa=0.5$, $\sigma=0$,
 $c=10^5$, and $\hat{c}=10^2$.
 
 Computations have been performed
 on uniform Cartesian grids
 of $2560^2$
 and $200^3$ zones in 2D and 3D respectively, for a total time
 $t=0.007$. 
 Final
 $\mathrm{log}_{10}E_r$, $\mathrm{log}_{10}\rho$, $\vert\vert\mathbf{v}\vert\vert$
 and $f$ profiles in the 3D test
 are shown in Fig. \ref{fig:blast}
 at $z=0$. Two radiation fronts
 can be identified in the $f$ profile: an
 outer front, caused by the
 initial relaxation of the system,
 and an inner front, corresponding
 to the radiative diffusion from the overpressurized region.
 Matter is isotropically
 accelerated, reaching at that
 time a maximum velocity of
 $\vert\vert\mathbf{v}\vert\vert
 =7.27$ in the outer boundaries
 of the central region.

\begin{figure}[t]
\centering
\includegraphics[width=0.47\textwidth]
{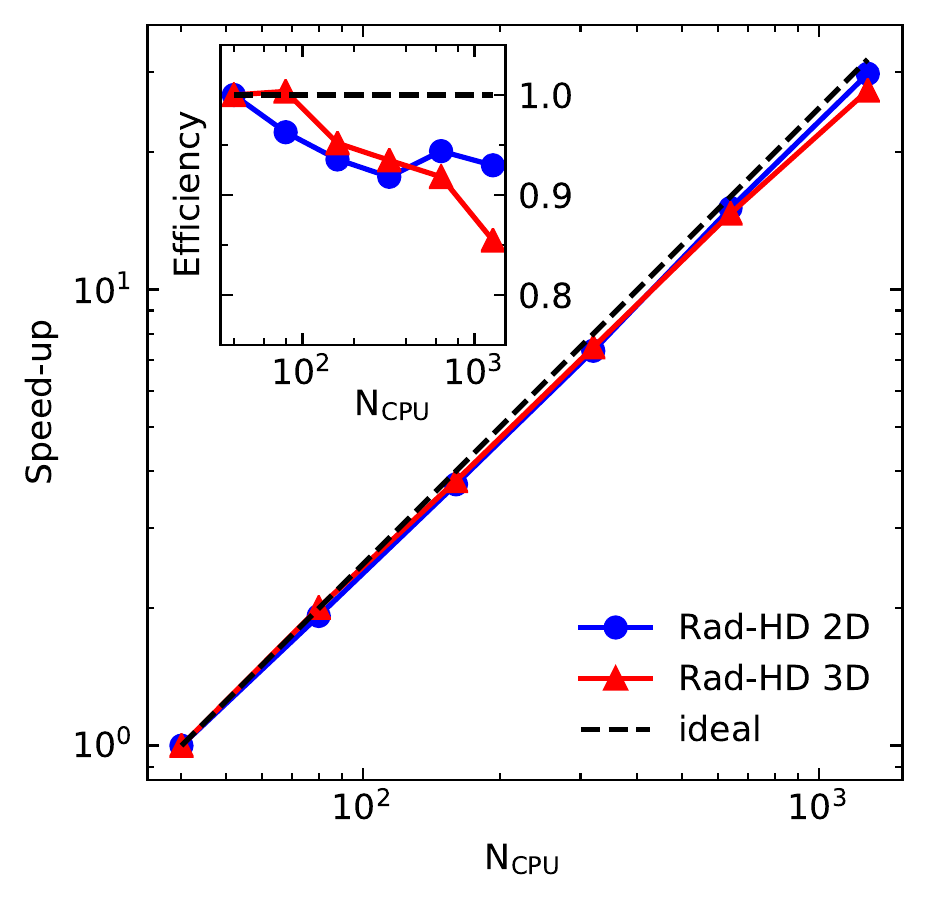}
\caption{ Speed-up factor
and scaling efficiency
for the 2D (blue) and 3D (red) blast-wave tests as a function of the number of processors. The ideal scaling law (dashed black line) is shown for comparison.}
\label{fig:scaling}
\end{figure}

 We ran each test using a different
 number of processors 
 (Intel Skylake 6148 at 2.2 GHz),
 varying from $N_\mathrm{CPU}=40$ to $1280$.
 We increased $N_\mathrm{CPU}$ in steps of $40$ given the $40$ cores per node
 architecture of our system. 
 Corresponding speed-up factors $S$
 are shown in Fig. \ref{fig:scaling} as a function of $N_\mathrm{CPU}$,
 computed as $S=T_\mathrm{ref}/T_{N_\mathrm{CPU}}$,
 where $T_{N_\mathrm{CPU}}$ is
 the average computation
 time per step
 for each $N_\mathrm{CPU}$,
 and $T_\mathrm{ref}=T_{40}$.
 In the same figure we show the
 obtained efficiencies for both the
 2D and 3D runs, all of which
 stay above $90\%$ for
 $N_\mathrm{CPU}\leq 512$,
 reaching $93\%$ and $85\%$ for
 $N_\mathrm{CPU}=1280$
 in 2D and 3D respectively.
 This scaling behavior is essential
 to overcome the scale disparity between
 radiation and HD  characteristic speeds,
 which makes Rad-HD computations
 approximately $120$ times
 more expensive than HD runs of
 this test. Some factors that in general affect
 the scaling efficiency of the code
 are the chosen
 domain decomposition, the
 latency that can
 arise if the condition
 $\vert\vert \mathbf{F}_r \vert\vert
 \leq E_r$ is imposed in only part of the domain,
 and the increasing number of
 communications for larger
 $N_\mathrm{CPU}$ required, e.g.,
 to define field values at ghost cells
 and to compute the time step.

